\documentclass[aps, prx, superscriptaddress, twocolumn, longbibliography]{revtex4-1}

\usepackage{graphicx}
\usepackage{amsmath}
\usepackage{amssymb}
\usepackage{hyperref}
\hypersetup{
 pdfnewwindow=true,
 colorlinks=true,
 linkcolor=blue,
 citecolor=blue,
 filecolor=blue,
 urlcolor=blue
}
\usepackage{color}
\usepackage{psfrag}
\usepackage{soul}
\usepackage{textcomp}
\usepackage{bbm}

\DeclareMathOperator{\tr}{Tr}

\DeclareMathOperator*\lowlim{\underline{lim}}
\DeclareMathOperator*\uplim{\overline{lim}}

\newcommand{\bea}{\begin{eqnarray}}
\newcommand{\eea}{\end{eqnarray}}
\newcommand{\besa}{\begin{subequations}\begin{eqnarray}}
\newcommand{\eesa}{\end{eqnarray} \end{subequations}}
\newcommand{\beaa}{\begin{eqnarray}\begin{aligned}}
\newcommand{\eeaa}{\end{aligned}\end{eqnarray}}

\newcommand{\comments}[1]{}
\newcommand{\av}[1]{\left\langle #1 \right\rangle}
\newcommand{\ket}[1]{| #1 \rangle}
\newcommand{\bra}[1]{\langle #1 |}
\newcommand{\ci}{\mathrm{i}}
\newcommand{\id}{\mathbbm{I}}
\newcommand{\p}[1]{\left( #1 \right)}

\newcommand{\gT}{\mathfrak{T}}
\newcommand{\CL}{\mathrm{(CL)}}
\newcommand{\tom}{\widetilde{\omega}}
\newcommand{\lom}{\overline{\omega}}
\newcommand{\cd}{\mathrm{c:d}}

\begin{document}

\title{Charging assisted by thermalization}

\author{Karen V. Hovhannisyan}
\affiliation{The Abdus Salam International Centre for Theoretical Physics (ICTP), 34151 Trieste, Italy}
\affiliation{A. Alikhanyan National Science Laboratory (Yerevan Physics Institute), 0036 Yerevan, Armenia}

\author{Felipe Barra}
\affiliation{Departamento de F\'{i}sica, Facultad de Ciencias F\'{i}sicas y Matem\'{a}ticas, Universidad de Chile, 837.0415 Santiago, Chile}

\author{Alberto Imparato}
\affiliation{Department of Physics and Astronomy, Aarhus University, 8000 Aarhus, Denmark}

\begin{abstract}

A system in thermal equilibrium with a bath will generally be in an athermal state, if the system-bath coupling is strong. In some cases, it will be possible to extract work from that athermal state, after disconnecting the system from the bath. We use this observation to devise a battery charging and storing unit, simply consisting of a system, acting as the battery, and a bath. The charging cycle---connect, let thermalize, disconnect, extract work---requires very little external control and the charged state of the battery, being a part of global thermal equilibrium, can be maintained indefinitely and for free. The efficiency, defined as the ratio of the extractable work stored in the battery and the total work spent on connecting and disconnecting, is always $\leq 1$, which is a manifestation of the second law of thermodynamics. Moreover, coupling, being a resource for the device, is also a source of dissipation: the entropy production per charging cycle is always significant, strongly limiting the efficiency in all coupling strength regimes. We show that our general results also hold for generic microcanonical baths. We illustrate our theory on the Caldeira-Leggett model with a harmonic oscillator (the battery) coupled to a harmonic bath, for which we derive general asymptotic formulas in both weak and ultrastrong coupling regimes, for arbitrary Ohmic spectral densities. We show that the efficiency can be increased by connecting several copies of the battery to the bath. Finally, as a side result, we derive a general formula for Gaussian ergotropy, that is, the maximal work extractable by Gaussian unitary operations from Gaussian states of multipartite continuous-variable systems.

\end{abstract}

\maketitle

\section{Introduction}

The second law of thermodynamics, as per the Kelvin-Planck formulation \cite{Fermi}, states that no work can be extracted in a cyclic manner from a system in thermal equilibrium. On the other hand, in the presence of strong interactions, the reduced state of a subsystem of a thermal system will not typically be thermal \cite{Ford_1965, Haake_1985}. Thereby, not limited by the Second Law anymore, it may be possible to cyclically extract nonzero work from such a subsystem, when manipulated in separation from the rest of the system \cite{Allahverdyan_2000, Nieuwenhuizen_2002}.

In precise terms, the cyclic processes referred to above are processes where the system's state evolves according to an externally driven time-dependent Hamiltonian, the value of which at the end is equal to that at the beginning. The maximal amount of work extractable from a system by such processes is called ergotropy \cite{Allahverdyan_2004} (see Appendix~\ref{app:ergdef} for a detailed definition). A system with zero ergotropy is called passive \cite{Pusz_1978, Lenard_1978}, and that with nonzero ergotropy is called active. In these terms, the previous paragraph reads: Thermal states are passive; however, the reduced state of a subsystem of a thermal system can be active with respect to the local Hamiltonian.

Inspired by these basic observations and the setup of Ref.~\cite{Barra_2019}, we introduce a battery charging cycle, the central idea of which is to connect a system (the ``battery'') in a passive (``depleted'') state to a thermal bath, and wait until they thermalize. This will prepare the system in an active (``charged'') state, from which we will be able to extract work \textit{after} the system is disconnected from the bath. Of course, this cycle does not violate the Kelvin-Planck formulation of the Second Law since connecting and disconnecting the system will cost work, which will have to be provided by external agents \cite{Allahverdyan_2000, Nieuwenhuizen_2002, Ford_2006}.

Since large thermal baths generically thermalize finite-size systems that come in contact with them (see the discussion in Sec.~\ref{sec:engine} and Appendix~\ref{app:thermalization}), our device offers two main advantages: (i) the creation of the battery's charged state requires no fine external control and, since thermalization is the process preparing that state, is robust against minor variations of the system-bath interaction; (ii) it costs nothing to maintain the charged state for as long as might be needed---it is the stationary state of system-bath interaction. These advantages are not simultaneously met in other battery designs which either require finely tuned external fields to perform unitary charging operations on the depleted state of the battery and assume that the battery is isolated after it is charged \cite{Campaioli_2018, Santos_2019}; or require system-bath interaction engineering \cite{Barra_2019}; or, in order to prevent the battery from leaking the charge, either rely on
fragile symmetries of the system-bath interaction \cite{Liu_2019, Hovhannisyan_2019} or actively manipulate the battery \cite{Pirmoradian_2019, Gherardini_2020, Kamin_2019}.


Although, at the beginning of the cycle, the state of the total system is not thermal---especially when the bath is not in a canonical (a.k.a. Gibbs) state, e.g., when it is in a microcanonical state---and therefore may be active, we show for a generic thermalizing bath that, due to cyclicity, the total work one has to spend on connecting and disconnecting the system from the bath, $W_\cd$, is always larger than the maximal work one can extract from the system after it is detached from the bath, i.e., its ergotropy $\mathcal{E}$. This means that the efficiency, defined as the ratio of the energy one is able to extract and the energy one has to invest for that: $\eta := \mathcal{E}/W_\cd$, is $\leq 1$. We show that this is nevertheless a consequence of the passivity of Gibbs states, even in those cases when the bath is microcanonical, and hence active \cite{Allahverdyan_2011}, due to the so-called equivalence of canonical and microcanonical states \cite{Riera_2012, Muller_2015, Brandao_2015}.



We support our general findings with a detailed calculation of all relevant quantities for the Caldeira-Leggett model \cite{Weiss_1999}, where the system is a harmonic oscillator and couples to a bath made of harmonic oscillators. We study the operation of the device in all relevant parameter regimes. In particular, we show that, other parameters fixed, the device is most efficient in the intermediate range of couplings. On the route of exploring different parameter regimes, we derive exact asymptotic expansions for the covariance matrix of the oscillator---for general Ohmic spectral densities---in the weak-coupling and ultrastrong-coupling. In the high-temperature limit, we prove an equipartition result for both the kinetic and potential energies of the oscillator, for arbitrary spectral densities and strengths of coupling. Found general expansions can be useful beyond the type of problems studied in this work.

Let us note in passing that, taking a different standpoint, our device can be viewed as a single-bath machine, and our above definition of its efficiency is standardly used in other types of single-bath machines \cite{Bustamante_2001, Seifert_2011, Golubeva_2012, Golubeva_2012a, Imparato_2015, Sune_2019, Barra_2019}. Importantly, however, the working principle of our device is fundamentally different from that of other single-bath machines such as molecular motors or force-to-force converters \cite{Bustamante_2001, Seifert_2011, Golubeva_2012, Golubeva_2012a, Imparato_2015, Sune_2019} in that, as opposed to those machines, in our case, the thermalizing effect of the environment has a constructive role in generating the output work. More generally, our device utilizes an energy storage mechanism which is not characteristic of the conventional molecular motors and force-to-force converters.

\section{The cycle}
\label{sec:engine}

Let us formalize the description of the device and its charging cycle in the introduction. First, we introduce the total system-bath Hamiltonian
\bea
H_\gT = H_s + H_B + H_I = H_0 + H_I,
\eea
where $H_s$ is the system Hamiltonian, $H_B$ is the bath Hamiltonian, $H_I$ is the interaction Hamiltonian (possibly containing a system renormalization term), and $H_0 = H_s + H_B$ is the bare, noninteracting total Hamiltonian of the system and the bath. We assume the bath to be large, i.e., consisting of a large $N \gg 1$ number of, generically, interacting constituents, and ``complex'' enough to thermalize nonmacroscopic systems that come in contact with it (see below for precise definitions). We denote the state of the system before interacting with the bath by $\rho_0$ and the initial state of the bath by $R_B$, so that the initial state of the total system is
\bea \label{inista}
\Omega = \rho_0 \otimes R_B.
\eea
Also, we introduce $\Omega_\infty = e^{- \ci H_\gT t_\infty} \Omega e^{\ci H_\gT t_\infty}$, where $t_\infty$ is a duration long enough to equilibrate the overall system. With these, the (Hamiltonian) cycle we consider consists of the following strokes:
\besa \label{step1}
&\fbox{1}& \hspace{2.5mm} \big[ H_0, \Omega \big] \overset{-W_c}{\xrightarrow{\hspace*{7mm}}} \big[ H_\gT, \Omega \big]
\\ \label{step2}
&\fbox{2}& \hspace{2.5mm} \big[ H_\gT, \Omega \big] \xrightarrow{\hspace*{7mm}} \big[ H_\gT, \Omega_\infty \big],
\\ \label{step3}
&\fbox{3}& \hspace{2.5mm} \big[ H_\gT, \Omega_\infty \big] \overset{- W_d}{\xrightarrow{\hspace*{7mm}}} \big[ H_0, \Omega_\infty \big],
\\ \label{step4}
&\fbox{4}& \hspace{2.5mm} \big[ H_0, \Omega_\infty \big] \overset{\mathcal{E}}{\xrightarrow{\hspace*{7mm}}} \big[ H_0, U_{\mathrm{erg}} \Omega_\infty U_{\mathrm{erg}}^\dagger \big],
\\ \label{step1p}
&\fbox{1$\hspace{-0.2mm}'$\hspace{-0.7mm}}& \hspace{2.5mm} \big[\text{Discard old bath, attach new bath,}
\\ \nonumber
&\phantom{\fbox{1}}& \hspace{5.5mm} \text{start the next cycle from} \, \rho_p \otimes R_B. \big] ~~~~~
\eesa
Here, the unitary operator
\bea
U_{\mathrm{erg}} = U_{\mathcal{E}} \otimes \id_B
\eea
acts solely on the system, and is such that $U_{\mathcal{E}}$ extracts the full ergotropy from $\rho_s^\infty = \tr_B \Omega_\infty$, with respect to $H_s$, and leaves the system in the passive state
\bea \label{ahuacatl}
\rho_p = \tr_B \big( U_{\mathrm{erg}} \Omega_\infty U_{\mathrm{erg}}^\dagger \big),
\eea
which is the state in which the system enters the next cycle. (See Appendix~\ref{app:ergdef} for precise definitions of ergotropy and passivity.)

As we mentioned, connecting and disconnecting the
system to the bath has a work cost $W_\cd := W_c + W_d$,
where $W_c$ is the cost of stroke \fbox{1} and $W_d$ of stroke \fbox{3}.

Remarkably, the stroke \fbox{2} can last as long as one wishes. Being the stage at which the active state of the system is prepared, it gives one considerable flexibility in practical situations, as there is no need of fine control---one just leaves the system in contact with a bath, and, whenever work is needed, one abruptly detaches the system (stroke \fbox{3}) and performs the extraction process (stroke \fbox{4}), which, when the bath is thermalizing (see below), is conveniently independent of the initial state of the system. Moreover, the time interval between \fbox{3} and \fbox{4} is also arbitrary, since the ergotropy of a system evolving under the influence of its own Hamiltonian is time-independent (see Appendix~\ref{app:ergdef}). However, the unitary operation extracting the ergotropy does depend on time, so, although one does not need to perform \fbox{4} immediately after \fbox{3}, fine tuning is necessary in any case. Importantly, in contrast to the lack of upper bound on the duration of stroke \fbox{2}, there is a lower bound: the equilibration time. Although it can sometimes be quite short \cite{Linden_2009, Goldstein_2013, Garcia-Pintos_2017}, relaxation generically takes nonnegligible amount of time which depends on the details of the system-reservoir interaction and the initial state. This time---the charging time in our case---is generally expected to be a decreasing function of the coupling strength \cite{Weiss_1999, Garcia-Pintos_2017, Perarnau-Llobet_2018}, especially in those situations when the initial state commutes with the bare Hamiltonian \cite{Garcia-Pintos_2017}, which is what we will typically have in our protocol. Indeed, starting from the second cycle, the initial state is $\rho_p \otimes R_B$, and passive states always commute with the Hamiltonians with respect to which they are passive ($[\rho_p, H_s] = 0$; see \cite{Pusz_1978, Lenard_1978, Allahverdyan_2004} or Appendix~\ref{app:ergdef}). Moreover, $R_B$ will generically---although not necessarily---be either a canonical or a microcanonical state, and both commute with $H_B$ (cf. Eqs.~\eqref{eq:tauB} and \eqref{microcan} below). This is a fortunate situation for our device: the need for strong (but not too strong---see Sec.~\ref{sec:CLenergetics}) coupling in order to operate comes with the benefit that it also ensures fast charging.

In our further analysis, we will assume---and this is a crucial assumption---that the joint evolution of the system and bath is thermalizing in the following sense: Say, $\mathfrak{S}$ is a finite subsystem of the system-plus-bath composite belonging to $s \cup \mathrm{supp}(H_I)$, then the long-time limit of the reduced state of $\mathfrak{S}$ should be
\bea \label{thermaliz}
\rho^\infty_{\mathfrak{S}} := \tr_{\gT \backslash \mathfrak{S}} \Omega_\infty = \tr_{\gT \backslash \mathfrak{S}} \tau_\gT,
\eea
where
\bea \label{taub}
\tau_{\gT} := \frac{e^{-\beta H_\gT}}{Z_\gT}.
\eea
More precisely, Eq.~\eqref{thermaliz} should be understood as a statement about the long-time average of the state:
\bea \label{avtherm}
\lim\limits_{\mathcal{T} \to \infty} \frac{1}{\mathcal{T}} \int_0^\mathcal{T} dt \Vert \tr_{\gT \backslash \mathfrak{S}} \Omega_t - \tr_{\gT \backslash \mathfrak{S}} \tau_\gT \Vert,
\eea
where $\Omega_t = e^{- \ci H_\gT t} \Omega e^{\ci H_\gT t}$ and $\Vert \cdot \Vert$ is the trace norm \cite{mikeike}, should be small and tend to zero as the size of the bath ($N$) increases (see Refs.~\cite{Linden_2009, Short_2012, Farrelly_2017, Garcia-Pintos_2017} for examples of such bounds). By the Markov inequality, this means that
the system's state will be close to $\tr_{\gT \backslash \mathfrak{S}} \tau_\gT$ for most of the time.

This is a rather weak assumption as thermalization is ubiquitous in macroscopic systems \cite{Robinson_1973, Bach_2000, Reimann_2008, Linden_2009, Short_2012, Riera_2012, Goldstein_2013, Muller_2015, Brandao_2015, Gogolin_2016, Farrelly_2017}. When the bath starts in a canonical (Gibbs) state,
\bea \label{eq:tauB}
R_B = \tau_B := \frac{e^{-\beta H_B}}{Z_B},
\eea
thermalization in the above sense is sometimes referred to as ``return to equilibrium.'' It has been rigorously proven in several generic scenarios. First is when, to an infinitely large, strictly continuous bosonic bath, is linearly coupled a continuous-variable system (e.g., the Caldeira-Leggett model \cite{Weiss_1999}) \cite{Mori_2008, Subasi_2012} or a system with a finite-dimensional Hilbert space (e.g., the spin-boson model \cite{Leggett_1987}) \cite{Bach_2000, Konenberg_2016}. The other scenario is when the bath is a many-body system with short-range interactions, away from criticality (i.e., with a finite correlation length), satisfying some transport conditions, e.g., nonzero Lieb-Robinson velocity (``speed of sound'') \cite{Nachtergaele_2010} or absence of many-body localization \cite{Gogolin_2016}. Note that exponential decay of correlations is very common in short-range interacting systems and is guaranteed for arbitrary Fermi systems at nonzero temperature \cite{Hastings_2004F}, vacuum states of gapped lattice Hamiltonians \cite{Hastings_2006}, general lattice systems above a critical temperature \cite{Kliesch_2014}, and is often related to the presence of a finite Lieb-Robinson velocity in the system \cite{Fredenhagen_1985, Hastings_2004, Nachtergaele_2006}. Early results about thermalization in systems with short-range interactions deal with the thermalization of locally perturbed translationally invariant infinite chains satisfying certain ergodicity-like properties (see, e.g., \cite{Robinson_1973} and references therein). More recent results for lattice systems are based on the ideas of equilibration \cite{Linden_2009, Short_2012, Garcia-Pintos_2017} and the equivalence of canonical and microcanonical ensembles and Berry-Esseen--type concentration bounds \cite{Riera_2012, Muller_2015, Brandao_2015, Farrelly_2017, Tasaki_2018}, and therefore require only few generic assumptions in order to hold. The most general rigorous proof of thermalization in the sense of Eq.~\eqref{thermaliz} holds for any finite-range $H_{\gT}$ (i.e., containing at most $k$-body interaction terms, where $k$ is some finite number) such that both $\tau_{\gT}$ and $\tau_B$ have exponentially decaying correlations \cite{Farrelly_2017}, with the only additional (weak) requirement being that $H_\gT$ needs to have not too many repeating gaps in its spectrum (see Appendix~\ref{app:thermalization} for more details). Using the fact that, with these conditions, microcanonical and canonical states are locally equivalent \cite{Brandao_2015, Tasaki_2018}, in Appendix~\ref{app:thermalization}, and combining results from Refs.~\cite{Brandao_2015, Farrelly_2017}, we show that subsystems thermalize (i.e., Eq.~\eqref{thermaliz} holds) also when the bath starts in the microcanonical state \cite{ll5old, Muller_2015, Brandao_2015}:
\bea \label{microcan}
R_B = \mu_B(E, \Delta) := \frac{1}{d(E, \Delta)} \sum_{|e_n - E| \leq \Delta} \ket{e_n} \bra{e_n}, ~~~~
\eea
where $e_n$ and $\ket{e_n}$ are, respectively, the eigenvalues and eigenvectors of $H_B$, $O\big( \ln^{2d} N \big) \leq \Delta \leq O\big( \sqrt{N} \big)$ is the microcanonical energy window, $d(E, \Delta)$ is the amount of energy levels in the interval $[E - \Delta, E + \Delta]$, and $E$ (chosen to be macroscopic, i.e., $\propto N$) is the energy of the microcanonical state. Temperature is prescribed to the microcanonical state through $E = \tr(H_B \tau_B)$, which, in view of $\tr(H_B \tau_B)$ also being $\propto N$ and monotonic with respect to $\beta$, defines a unique function $\beta = \beta(E/N)$.

In view of the thermalization assumption, let us point out that, whereas, similarly to stroke \fbox{2}, stroke \fbox{1} can last as long as one wishes ($W_c$ will of course depend on the details of the switching protocol, but the end result of stroke \fbox{2} will always be the same), stroke \fbox{3} has to be a nonstationary process. Indeed, if performed in a slow, ``quasi-equilibrium'' manner, the system will at all times remain thermalized with the bath. Therefore, by the end of the disconnection process, when the interaction vanishes, it will simply be in a Gibbs state with respect to $H_s$, hence, no work will be possible to extract during stroke \fbox{4}, rendering the cycle useless.

We also note that, since the Ohmic Caldeira-Leggett model with a Lorentz-Drude cutoff function can be mapped into a gapless harmonic lattice (where the system maps to a node) with polynomially decaying interactions \cite{Hovhannisyan_2018}, the thermalization result for the CL model \cite{Mori_2008, Subasi_2012} can be thought of as an extension of the above-described paradigm of [short-range]$+$[noncritical]$\rightarrow$[thermalization] to critical systems with long-range interactions; in this context, the result about the microcanonical bath also being thermalizing is unlikely to hold.

\subsection{The energetics of the cycles}

Here we will study the energetics of the device and find its efficiency. Directly reading from Eqs.~\eqref{step1}--\eqref{step4}:
\besa \label{ene1}
W_c &=& \tr[H_I \Omega],
\\ \label{ene2}
W_d &=& - \tr[H_I \Omega_\infty],
\\ \label{ene3}
\mathcal{E} &=& \tr[H_s \rho_s^\infty] - \tr[H_s \rho_p],
\eesa
where $\rho^\infty_s$ is given by Eq.~\eqref{thermaliz}. Mind the sign convention: $W_c$ and $W_d$ are \textit{works performed on} the (composite) system, whereas $\mathcal{E}$ is the \textit{work extracted from} the system.

Relying on the thermalization results discussed in the previous subsection, we will henceforth assume that the system-bath evolution is thermalizing for all finite subsystems of $s \cup \mathrm{supp}(H_I)$, and therefore, in Eqs.~\eqref{ene1} and \eqref{ene2}, we will substitute $R_B$ by $\tau_B$ (if $R_B$ is not $\tau_B$, e.g., when it is the microcanonical state \eqref{microcan}) and $\Omega_\infty$ by $\tau_\gT$:
\bea \label{ene1th}
W_\cd = W_c + W_d = \tr[H_I \rho_0 \otimes \tau_B] - \tr[H_I \tau_\gT]. ~~~~
\eea
Here we need to maintain caution, since, as is detailed in Appendix~\ref{app:thermalization}, Eq.~\eqref{thermaliz} generally holds only up to a correction $O\left( N^{-\epsilon} \right)$ (with some $\epsilon >0$) if $\mathrm{supp} (H_I)$ is finite. However, when $\mathrm{supp} (H_I)$ scales with $N$, the corrections may potentially accumulate into something nonnegligible. We discuss this further in Appendix~\ref{app:ThermInfSupp}, where we show the conditions on $H_I$ that guarantee the correctness of Eq.~\eqref{ene1th}; for the Caldeira-Leggett model, these reduce to an explicit condition on the spectral density.

Now, defining ``dissipated work'' as
\bea \label{dissdef}
W_\mathrm{diss} = W_\cd - \mathcal{E},
\eea
which, according to our sign convention, corresponds to
the total work performed during the cycle, and putting Eqs.~\eqref{ene1}--\eqref{ene1th} together, we find that
\bea \label{promezh1}
W_\mathrm{diss} = w^{(1)} + T \Sigma,
\eea
where
\bea \label{zhar1}
w^{(1)} = \tr\big[H_I \big(\rho_0 - \rho_p \big) \otimes \tau_B\big]
\eea
and
\beaa \label{wombat}
T \Sigma = & \, \tr[H_I \rho_p \otimes \tau_B] - \tr[H_I \tau_\gT] 
\\
& \, + \tr[H_s \rho_p] - \tr[H_s \rho_s^\infty];
\eeaa
the reason for this notation will become clear in what follows.

Starting from the second cycle, the information about the initial state of the system, $\rho_0$, is lost: the system starts and ends in the state $\rho_p$ (see Eqs.~\eqref{step1p} and \eqref{ahuacatl}), meaning that
\bea
w^{(i)} = 0, \quad \text{for} \quad i \geq 2,
\eea
where $i$ counts the cycles. In fact, in most physically relevant situations, $w^{(1)}$ will also be $= 0$. Indeed, generically, $H_I = \sum_\kappa g_\kappa S_\kappa \otimes B_\kappa$, where $S_\kappa$ are some system operators and $B_\kappa$ are bath operators. The latter will typically be some sort of ``field operators'', and, above a critical temperature (and at any temperature in one- and two-dimensional systems with continuous symmetries), will generically have zero thermal averages: $\tr[B_\kappa \tau_B] = 0$ (see the discussion of the Mermin-Wagner theorem in, e.g., \cite{Friedli_2017}), meaning that $w^{(1)} = 0$ irrespective of $\rho_0$. This is obviously the case for all quadratic bosonic \cite{Leggett_1987, Weiss_1999} and fermionic baths \cite{Lieb_1961}. That said, note that we need not and will not make such an assumption in what follows. In fact, we could get rid of $w^{(1)}$ altogether, by permuting the strokes of our cycle in Eqs.~\eqref{step1}--\eqref{step1p}: \fbox{2} $\to$ \fbox{3} $\to$ \fbox{4} $\to$ \fbox{1$\hspace{-0.2mm}'$\hspace{-0.6mm}} $\to$ \fbox{1}. For this ``reshuffled'' cycle, the initial state of the system will never enter the energetics, since the first cycle in this protocol will be equivalent to the second cycle in the original protocol.

Now, let us come back to $T \Sigma$. Noting that $\tr[H_s \rho_p] = \tr[H_s \otimes \id_B \rho_p \otimes \tau_B]$ and $\tr(H_s \rho_s^\infty) = \tr(H_s \otimes \id_B \tau_\gT)$, and introducing $\rho_B^\infty = \tr_s \tau_\gT$, we can rewrite Eq.~\eqref{wombat} as
\beaa \label{promezh2}
T \Sigma = & \, \tr[H_\gT \rho_p \otimes \tau_B] - \tr[H_\gT \tau_\gT]
\\
& \, - \tr[H_B \tau_B] + \tr[H_B \rho_B^\infty].
\eeaa
Note that this formula holds both when $R_B = \tau_B$ and $R_B = \mu_B(E, \Delta)$ (and whenever the bath is thermalizing).

Furthermore, introducing
\bea \label{promezh3}
\tau_\gT' = \frac{e^{-\beta' H_\gT}}{Z_\gT'},
\eea
where $\beta'$ is determined from
\bea \label{promezh4}
S(\tau_\gT') = S(\rho_p \otimes \tau_B) = S(\rho_s^\infty) + S(\tau_B),
\eea
where $S(\rho) = - \tr(\rho \ln \rho)$ is the von Neumann entropy and, in the second equality, we noted that $\rho_p$ and $\rho_s^\infty$ have the same entropy because one is a unitary transformation of the other. Next, by a series of algebraic manipulations, presented in Appendix~\ref{app:EPderivation}, we prove the following identity:
\beaa \label{final1}
\Sigma = & \, \frac{T'}{T} S(\rho_p \otimes \tau_B \Vert \tau_\gT') + S (\tau_\gT' \Vert \tau_\gT)
\\
& \, + I_{\tau_\gT}(s : B) + S(\rho_B^\infty \Vert \tau_B),
\eeaa
where $S(\rho || \tau) = \tr [\rho (\ln \rho - \ln \tau)]$ is the relative entropy and $I_{\tau_\gT}(s : B) = S(\rho_s^\infty) + S(\rho_B^\infty) - S(\tau_\gT) \geq 0$ is the mutual information \cite{mikeike} between the system and the bath in the state $\tau_\gT$.

All the terms in the RHS of Eq.~\eqref{final1} are nonnegative, therefore,
\bea \label{final2}
\Sigma \geq 0,
\eea
which, given that $T \Sigma$ is the dissipated work, allows us to loosely interpret $\Sigma \geq 0$ as the entropy production of the cycle \cite{Esposito_2010}. In the following subsection, we will corroborate this interpretation by putting the energetics of the cycle in the context of the Second Law.

\subsection{Discussion of the energetics}
\label{sec:discE}

Ergotropy, $\mathcal{E}$, is by definition nonnegative, and since $T \Sigma \geq 0$, the connection-disconnection work will also be nonnegative:
\bea \label{workin1}
W_\cd \geq 0.
\eea
In those cases when $w^{(1)}$ happens to be nonzero, it can be both positive and negative, therefore, (only) for the first cycle, $W_\cd \geq 0$ may not hold.

Therefore, in the general spirit of thermodynamics, and particularly the example of other single-bath machines \cite{Bustamante_2001, Seifert_2011, Barra_2019}, we will define the efficiency of our device as an output/expenditure ratio:
\bea \label{effic1}
\eta := \frac{\mathcal{E}}{W_\cd} = 1 - \frac{T \Sigma}{W_\cd} \leq 1.
\eea

System-bath coupling is what powers our device, which is expressed in the fact that
\bea \label{philos1}
T\Sigma = \mathcal{E} = 0, \quad \text{whenever} \quad H_I = 0,
\eea
as follows from Eqs.~\eqref{ene1}--\eqref{ene3}. Suppose for a moment that there exists a single parameter, $g$, that controls the strength of the interaction (say, $H_I = g V$). Then, assume that $\Sigma$ is differentiable in $g$ (this is not guaranteed as $U_\mathcal{E}$ is a permutation matrix that depends on eigenvalue ordering). Now, since both the relative entropy and mutual information are nonnegative, their Taylor expansion for small $g$'s cannot start with a $\propto g$ term, because otherwise the $g \to - g$ transformation would change the sign of the quantity, for sufficiently small $g$'s; therefore, $T \Sigma = O \big(g^{2 k}\big)$, where $k \geq 1$ is a natural number. For the same reason, since $W_\cd \geq 0$ and $\to 0$ as $g \to 0$, we also conclude that $W_\cd = O \big(g^{2 m}\big)$, where $m \geq 1$ is another natural number (see Ref.~\cite{Pozas-Kerstjens_2018} for an example of an explicit calculation of connection-disconnection work in a harmonic chain, with continuous switching, where $m = 1$). This means that, although both the nominator and denominator in Eq.~\eqref{effic1} go to zero, their ratio, and therefore $\eta$, will either go to zero (if $k > m$), or remain finite but $< 1$ (if $k = m$), or go to one (if $k < m$). The latter case is obviously the most interesting, however, we could not find such an example.

Now we will link Eq.~\eqref{final2} to the Second Law. With this purpose, we note that, by virtue of
Eqs.~\eqref{ene1th}--\eqref{promezh1}, the total work performed on the total system in $n$ cycles is $n T \Sigma$ (when $w^{(1)} \neq 0$, we need to add it as well, however, it will play no role for sufficiently large $n$'s, therefore, we will omit it). By definition, $- n T \Sigma$ is the net extracted work in $n$ cycles. On the other hand, when the bath is canonical, the bird-eye view on the cycle reveals a simple picture: the total system starts with $(\rho_0 \otimes \tau^{(n)}_B, H_s + H_B^{(n)})$, where $H_B^{(n)} = H_B + \cdots + H_B$ and $\tau_B^{(n)} = \tau_B^{\otimes n} \propto e^{- \beta H_B^{(n)}}$ (the subsequent argumentation does not rely on the baths being identical; we made that choice to merely simplify notation). Then, the overall Hamiltonian, $H_s + H^{(n)}_B$, undergoes a cyclic variation in time, thereby driving the overall system unitarily, and, at the end of the cycle, we are left with $(U \rho_0 \otimes \tau_B^{(n)} U^\dagger, H_s + H_B^{(n)})$, where $U$ is the unitary evolution operator generated by the cyclic variation of the Hamiltonian. Now, a well-known formulation of the Second Law (which was first noted in phenomenological thermodynamics in Refs.~\cite{Fermi, ll5old} and rigorously proven in the quantum regime in Ref.~\cite{Lindblad_1983}), states that the maximal work that can be extracted by a cyclic variation of the Hamiltonian (that is, ergotropy) from $(\rho_0 \otimes \tau, H_s + H)$ (where $H$ is an arbitrary Hamiltonian and $\tau \propto e^{- \beta H}$) is upper-bounded by the difference of nonequilibrium free energies of the system:
\bea \label{wmax}
W_{\max} \! \leq \! F_\beta \big[ H_s, \rho_0 \big] \! - F_\beta \big[H_s, \tau_s \big] \! = T S \big( \rho_0 \big\Vert \tau_s \big), ~~~~
\eea
where $\tau_s \propto e^{- \beta H_s}$ and $F_\beta [H, \rho] = \av{H} - T S = \tr[\rho H] + T \tr[\rho \ln \rho]$ is the nonequilibrium free energy of the system with respect to a $T$-temperature bath. This fact is a consequence of the simple identity
\bea \label{work}
W = T S(\rho_0 \Vert \tau_s) - T S(U \rho_0 \otimes \tau U^\dagger \Vert \tau_{sB}),
\eea
where $U$ is the unitary evolution operator generated by the cyclic variation of the Hamiltonian, $\tau_{sB} \propto e^{-\beta (H_s + H)}$, and $W = \tr[(H_s + H) \rho_0 \otimes \tau] - \tr[(H_s + H) U \rho_0 \otimes \tau U^\dagger]$ is the average extracted work. For sufficiently large baths, the upper bound in Eq.~\eqref{wmax} is always reachable through a sequence of small quenches and thermalizations (namely, $S(U \rho_0 \otimes \tau U^\dagger \Vert \tau_{sB})$ can be made arbitrarily small; see, e.g., Ref.~\cite{Aberg_2013}). Note that, in view of Eq.~\eqref{work}, extracting $T S(\rho_0 \Vert \tau_s)$ amount of work necessarily leaves the system in the state $\tau_s$.

It follows from Eq.~\eqref{wmax} that, even in the presence of $n$ copies of the bath, the maximal work extractable from all the systems altogether is upper-bounded by a fixed quantity, $T S (\rho_0 \Vert \tau_s)$, meaning that, if the net extracted work in a single cycle would be positive, after sufficiently many cycles, one would surpass $W_{\max}$, which is impossible. For our cycle, this indeed means that $\Sigma \geq 0$, which we now derived as a consequence of the Second Law. Interestingly, Eq.~\eqref{wmax} also implies that, since at the beginning of each cycle the Hamiltonian is $H_0$ and the state is $\rho_p \otimes \tau_B$ (or $\rho_0 \otimes \tau_B$ for the first cycle), the ergotropy of the system-plus-bath is $T S(\rho_p \Vert \tau_s) > 0$ (if the bath is large). However, as we noted above, spending that resource in one cycle would leave the system in the state $\tau_s$, thereby trivializing all subsequent cycles.

Note that the argument about no positive net extracted work applies to an arbitrary cycle, not just the one defined by Eqs.~\eqref{step1}--\eqref{step4}. More formally, any [attach]-[operate]-[detach]-type cycle (after the first one) of a single-bath machine can be thought of as a unitary transformation $\rho_p \otimes \tau_B \to \Omega' = U \rho_p \otimes \tau_B U^\dagger$ such that $\tr_B \Omega' = \rho_p$ ($U$ is the unitary evolution operator generated by the cyclic variation of the Hamiltonian, which depends on the particularities of the cycle). Now, it immediately follows from $I_{\Omega'}(s : B) \geq 0$ that $S(\tr_s \Omega') \geq S(\tau_B)$. Therefore, for the invested (``dissipated'') work we have: $W_{\mathrm{diss}} = \tr[(H_s + H_B) \Omega'] - \tr[(H_s + H_B) \rho_p \otimes \tau_B] = \tr[H_B \tr_s \Omega'] - \tr[H_B \tau_B]$, which, using the $H_B = - T \ln \tau_B - T \ln Z_B$ identity, we rewrite as
\bea \label{cat}
\beta W_{\mathrm{diss}} \! = S(\tr_s \Omega') - S(\tau_B) + S(\tr_s \Omega' \Vert \tau_B) \geq 0. ~~~~
\eea
This relation of course applies to our cycle as well. The merit of Eq.~\eqref{final1} for canonical baths is in the nuance that, in general, Eq.~\eqref{thermaliz} applies only to small subsystems of the whole, which means that the final state of the bath, $\tr_s \Omega'$, does not coincide with $\tr_s \tau_\gT$, and accessing the state of the bath is usually an intractable problem. Whereas Eq.~\eqref{final1} does not require knowledge of $\tr_s \Omega'$.

The content of Eq.~\eqref{final1} is much more nontrivial when the baths are microcanonical. In that case, the baths themselves are active, in the sense that the ergotropy of the bath (or the system-plus-bath) can be $\propto \sqrt{N}$ \cite{Allahverdyan_2011}. Moreover, it is possible to extract $\propto \sqrt{N}$ work from them in a (global) process that is cyclic in terms of both the Hamiltonian and the state of the system, in other words, there exist [attach]-[operate]-[detach] cycles which extract $O \big(\sqrt{N}\big)$ amount of work on each run. In this context, Eq.~\eqref{final2} (and Eq.~\eqref{zhar2} below, in the worst-case scenario for the first cycle) pose very strong restrictions on the operation of the device. These restrictions are the price we pay for designing our cycle in such a way that the maintenance of the active state comes for free.

We observe, in passing, that Eq.~\eqref{wmax} also allows us to lower-bound $w^{(1)}$. Indeed, since one can extract at most $T S(\rho_0 \Vert \tau_s)$ net amount of work in the first cycle,
\bea \label{zhar2}
w^{(1)} \geq - T S(\rho_0 \Vert \tau_s).
\eea
Importantly, this bound holds also when the bath is microcanonical, since the expression for $w^{(1)}$, Eq.~\eqref{zhar1}, is the same for canonical and microcanonical baths. 

Remarkably, in the context of Eq.~\eqref{philos1}, Eq.~\eqref{final1} provides yet another nontrivial insight: on the one hand, the device needs nonzero coupling in order to function, on the other hand, nonzero coupling inevitably leads to dissipation. This conflict is an analogue of the power-efficiency trade-off in ordinary thermal machines (see, e.g., \cite{Sekimoto_2000, Allahverdyan_2013, Shiraishi_2016}).

Finally, let us comment on the necessity of attaching the system to a new bath at the beginning of each cycle. If one would have to literally keep many copies of the bath in store in order to operate the device, then the whole setup would have virtually no practical significance. However, we expect a generic many-body, short-range-interacting bath which couples to the system locally will, upon the start of each cycle, appear to the system as if it were new. The idea is that, as we argued above, for the bath to be thermalizing, it needs to have exponentially decaying correlations, which generically means that there is nonzero Lieb-Robinson velocity in the bath \cite{Fredenhagen_1985, Hastings_2004, Nachtergaele_2006}. Therefore, once the system is detached from the interaction site of the bath, the area around the interaction site, perturbed by the interaction with the system, will diffuse away from the site, carrying the system-bath correlations along with it. Thus, by the time the next interaction session starts, the interaction site will be only slightly perturbed, as if the bath were fresh. An example explicitly illustrating such a behavior, on the specific model of the bath being a linear chain of harmonic oscillators with nearest-neighbor interactions, was reported in Ref.~\cite{Pozas-Kerstjens_2018}. Of course, when dealing with finite-size baths, after a certain amount of cycles, the perturbations will travel back to the interaction site. The rigorous formalization and justification of the described picture and the study of finite-size-bath effects delineate a motivating class of problems for future studies, but are beyond the scope of the present work.

\section{Caldeira-Leggett model for the device}
\label{sec:CLmodel}

We will now illustrate the theory developed above on the example of a harmonic oscillator (the system) linearly coupled to a bath consisting of independent harmonic oscillators and starting in a Gibbs state. This system-bath model is widely used to study quantum Brownian motion and is known as the Caldeira-Leggett (CL) model \cite{Weiss_1999}. Its total Hamiltonian, $H^\CL_\gT$, is thus defined through
\beaa \label{CLham}
H^\CL_s &= \frac{1}{2} p^2 + \frac{1}{2} \omega_0^2 q^2,
\\
H^\CL_I &= \frac{1}{2} \omega_R^2 q^2 - q \sum_k g_k q_k,
\\
H^\CL_B &= \sum_k \left[ \frac{p_k^2}{2 m_k} + \frac{m_k \omega_k^2 q_k^2}{2} \right],
\eeaa
where the renormalization frequency of the system, $\omega_R$ is given by
\bea \label{omer}
\omega_R^2 = \sum_k \frac{g_k^2}{m_k \omega_k^2},
\eea
and ensures that the Hamiltonian is bounded from below.

In the continuous limit, namely, when the bath frequencies, $\omega_k$, are very close to each other ($\omega_{k+1} - \omega_k \ll \omega_0$) and range from near-zero values to those significantly larger than $\omega_0$ (see, e.g., Ref.~\cite{Hovhannisyan_2018} for a careful treatment of the discrete-continuous transition), there exists a unique steady state to which the oscillator evolves \cite{Subasi_2012}. Moreover, this state is Gaussian, since it is given by Eq.~\eqref{thermaliz} \cite{Subasi_2012} and $\tau_\gT^\CL$ is a Gaussian state (as $H_\gT^\CL$ is quadratic). This means that the oscillator's steady state can be fully described by the first moments, $\av{q}_\infty$, $\av{p}_\infty$, and second moments,
\bea \label{lobzik}
\sigma^\infty_{ij} = \frac{1}{2} \av{\{ x_i, x_j \}}_\infty,
\eea
where $x = (q, p)$ and
\bea \nonumber
\av{\bullet}_\infty := \tr \big[ \bullet \Omega_\infty^\CL \big]
\eea
is the infinite-time average and $\sigma^\infty_{ij}$ comprise the so-called covariance matrix of the oscillator in the steady state. The covariance matrix can be defined for any state (see Eq.~\eqref{covmatdef} in Appendix~\ref{app:GErgotropy}), and it fully determines the state if the state is Gaussian.

Writing the Heisenberg equations,
\bea \label{Heq}
\ddot{q} + (\omega_0^2 + \omega_R^2) q = \sum_k g_k q_k, \quad \ddot{q}_k + \omega_k^2 q_k = \frac{g_k}{m_k} q, ~~~
\eea
and introducing the so-called spectral density,
\bea \label{jorik}
J(\omega) = \frac{\pi}{2} \sum_k \frac{g_k^2}{m_k \omega_k} \delta(\omega - \omega_k),
\eea
one can show that (see, e.g., Ref.~\cite{Weiss_1999} or Appendix~\ref{app:textbook}), as long as $J(\omega) \neq 0$ for $\omega \neq 0$, $\av{q}_\infty = \av{p}_\infty = \sigma_{12}^\infty = 0$ and
\bea \label{sgmii}
\sigma_{ii}^\infty = \frac{1}{\pi} \int_0^\infty d\omega \frac{J(\omega) \omega^{2 (i - 1)}}{|\alpha(\omega)|^2} \coth \frac{\omega}{2 T},
\eea
where $i$ is either $1$ or $2$ and $\alpha(\omega) = \omega_0^2 - \omega^2 + \omega_R^2 - \chi(\omega)$, with $\chi(\omega) = \frac{1}{\pi} \mathcal{P} \int_{-\infty}^\infty d \omega' \frac{J(\omega')}{\omega' - \omega}$ (the symbol $\mathcal{P}$ in front of an integral means the Cauchy principal value; see Appendix~\ref{app:textbook} for details).

In this paper (except for Appendix~\ref{app:sigmahighT}), we will work with so-called Ohmic spectral densities, which are characterized by a linear scaling of $J(\omega)$ with respect to $\omega$, for small $\omega$'s \cite{Weiss_1999}. Also, since a given oscillator with frequency $\omega_0$ cannot couple to bath modes with much higher frequencies, $J(\omega)$ must be a decaying function for $\omega \gg \omega_0$. With these in mind, we write the spectral density as
\bea \label{J:def}
J(\omega) = \gamma \omega_0 \omega f(\omega/\omega_c),
\eea
where $\omega_c$ is the cutoff frequency and $f(z)$ is a well-behaved dimensionless ``cutoff'' function that decays for $z > 1$ (i.e., when $\omega$ is above the cutoff frequency $\omega_c$) and $f(0) > 0$. In this notation, taking into account Eq.~\eqref{jorik} and Eq.~\eqref{omer},
\bea \label{omerta}
\omega_R^2 = \frac{2}{\pi} \int_0^\infty d \omega \frac{J(\omega)}{\omega} = \frac{2 \gamma \omega_0 \omega_c}{\pi} \int_0^\infty dz f(z), ~~~
\eea
therefore, in order for $\omega_R$ to be finite (so that the model has a physical meaning), $f(z)$ should decay fast enough for the second integral in Eq.~\eqref{omerta} to be convergent. Here, $\gamma$ is a dimensionless constant defining the coupling strength. The most common choices for the cutoff function in the literature are the Lorentz-Drude,
\bea \label{fLor}
f^{\mathrm{(L)}}(z) = \frac{2}{1 + z^2},
\eea
and exponential, $f^{\mathrm{(exp)}}(z) = \pi e^{-z}$, cutoff functions \cite{Weiss_1999}.

Eq.~\eqref{sgmii} clearly shows that $\rho_s^{\infty \, \CL} \neq \tau^\CL_s$. Indeed, $\tau^\CL_s$, the thermal state of a free (i.e., not interacting with any other system) harmonic oscillator with frequency $\omega_0$ is characterized by
\beaa \label{freeosc}
\sigma_{11}^{\mathrm{(free)}} = \frac{1}{2 \omega_0} \coth \frac{\omega_0}{2 T}, \quad \; \sigma_{22}^{\mathrm{(free)}} = \frac{\omega_0}{2} \coth \frac{\omega_0}{2 T}, ~~~~
\eeaa
and only in the zero-coupling limit do $\sigma_{11}^\infty$ and $\sigma_{22}^\infty$ coincide with $\sigma_{11}^{\mathrm{(free)}}$ and $\sigma_{22}^{\mathrm{(free)}}$, i.e., $\rho_s^{\infty \, \CL} \to \tau^\CL_s$ (see Appendix~\ref{app:weaksigma}). We emphasize that Eq.~\eqref{thermaliz}, i.e., $\rho_s^{\infty \, \CL} = \tr_B \tau_\gT^\CL$, holds in all coupling regimes.

\subsection{Energetics of the single-oscillator device}
\label{sec:CLenergetics}

For the Caldeira-Leggett model described above, we can derive explicit expressions for all the relevant quantities: the ergotropy, $\mathcal{E}^\CL$ (Eq.~\eqref{ene3}), the connection work, $W^\CL_c$ (Eq.~\eqref{ene1}), and the disconnection work, $W^\CL_d$ (Eq.~\eqref{ene2}).

Starting with the ergotropy, we show in Appendix~\ref{app:Gerg1osc} that the ergotropy of the state $\rho_s^{\infty \, \CL}$ (the covariance matrix of which is $\sigma^\infty$ in Eq.~\eqref{sgmii}, with respect to $H_s^\CL$, is given by
\bea \label{ergo:simp}
\mathcal{E}^\CL = \frac{1}{2} \p{ \sqrt{\sigma_{22}^\infty} - \omega_0 \sqrt{\sigma_{11}^\infty}}^2,
\eea
and the extraction of the ergotropy leaves the system in the Gaussian state $\rho^\CL_p$ characterized by the covariance matrix
\bea \label{s1:eq}
\sigma_p = \sqrt{\sigma_{11}^\infty \sigma_{22}^\infty} \p{\begin{array}{cc} \omega_0^{-1} & 0 \\ 0 & \omega_0 \end{array}},
\eea
which is in fact a thermal state at inverse temperature $\beta_p = 2 \omega_0^{-1} \mathrm{arccoth} \p{2 \sqrt{\sigma_{11}^\infty \sigma_{22}^\infty}}$ (see Appendix~\ref{app:Gerg1osc}). From Eq.~\eqref{ergo:simp} we immediately see that the ergotropy vanishes if and only if the steady state of the system, $\rho^{\infty \, \CL}_s$, is characterized by energy equipartition, namely, the average kinetic and potential energies are equal: $\sigma^\infty_{22} = \omega_0^2 \sigma^\infty_{11}$. As we will discuss below, this occurs in both weak-coupling and high-temperature limits. In the following, we will assume that the device has already performed the first cycle, so that the state of the total system at the beginning of the cycle (stroke \fbox{1}) is
\bea \nonumber
\Omega^\CL = \rho^\CL_p \otimes \tau^\CL_B,
\eea
where we emphasize again that $\rho^\CL_p$ is the ``exhausted'' state, characterized by the covariance matrix \eqref{s1:eq}. With this premise, invoking the fact that $\tr \big[ q_k \tau_B^\CL \big] = 0$, $\forall k$, and taking into account the definitions \eqref{CLham} and \eqref{lobzik}, the connecting work, $W_c^\CL = \tr \big[ H^\CL_I \Omega^\CL \big]$ (cf. Eq.~\eqref{ene1}), will read
\bea \label{wc:eq}
W^\CL_c = \frac{1}{2} \omega_R^2 \, (\sigma_p)_{11} = \frac {\omega_R^2}{2\omega_0} \sqrt{\sigma_{11}^\infty \sigma_{22}^\infty},
\eea
where, in the second equality, we used Eq.~\eqref{s1:eq}.

For the disconnecting work, $W_d = - \tr \big[ H^\CL_I \Omega^\CL_\infty \big]$ (cf. Eq.~\eqref{ene2}), we have 
\bea \label{wd:eq}
W^\CL_d = \av{q \ddot{q}}_\infty + \bigg(\omega_0^2 + \frac{1}{2} \omega_R^2 \bigg) \sigma_{11}^\infty,
\eea
where we have used Eq.~\eqref{Heq} to get rid of the $q \sum_k g_k q_k$ term in $H^\CL_I$. As we show in Appendix~\ref{app:textbook}, $\av{\ddot{q} q}_\infty = - \sigma_{22}^\infty$, therefore, with Eq.~\eqref{wc:eq},
\bea \label{wcd:eq}
W^\CL_\cd \! = \frac{\omega_R^2}{2\omega_0} \sqrt{\sigma_{11}^\infty \sigma_{22}^\infty } + \bigg( \omega_0^2 + \frac{1}{2} \omega_R^2 \bigg) \sigma_{11}^\infty - \sigma_{22}^\infty. ~~~~
\eea

\begin{figure}[t!]
\center
\includegraphics[width = 8.5 cm]{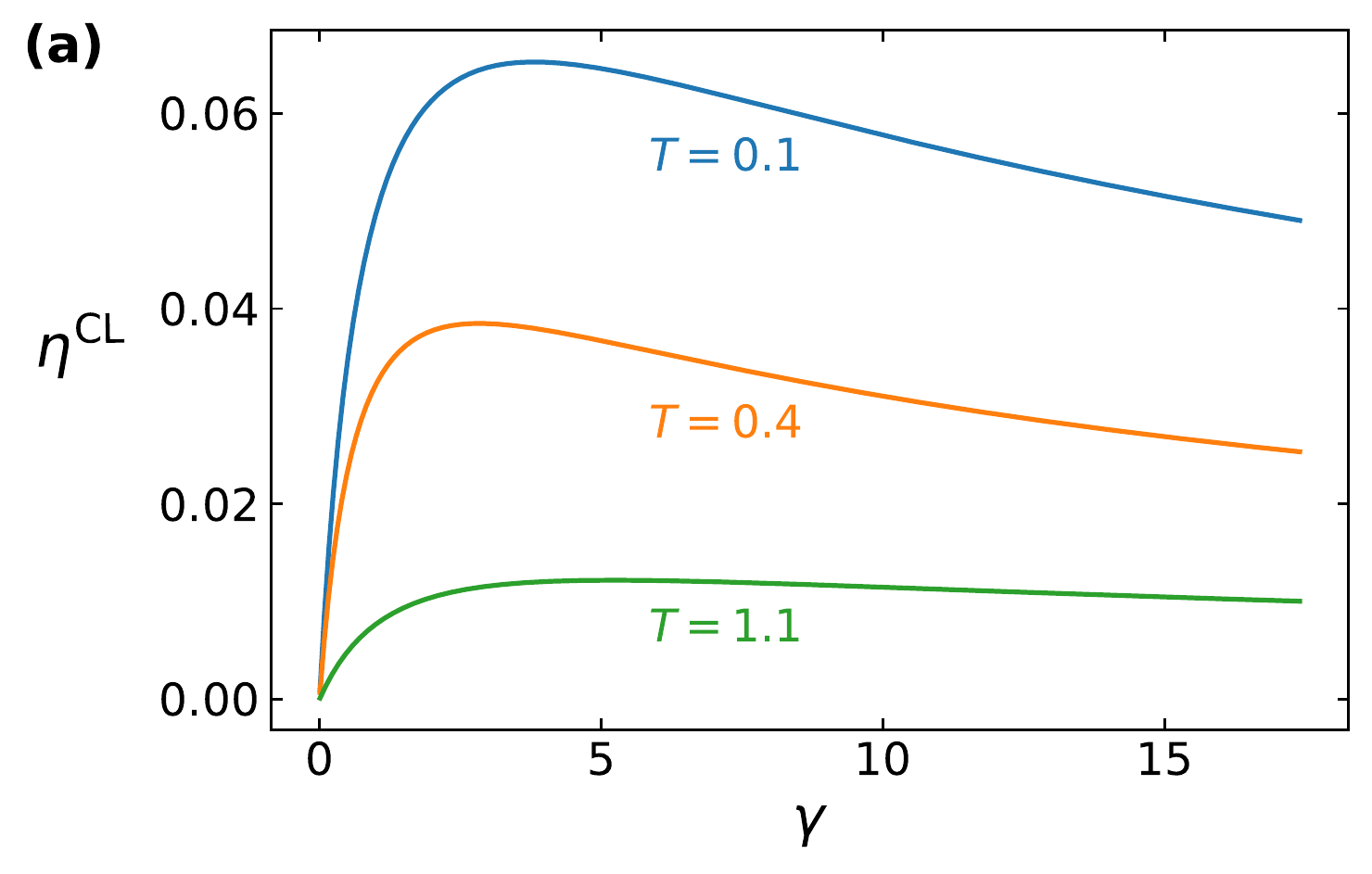}
\includegraphics[width = 8.5 cm]{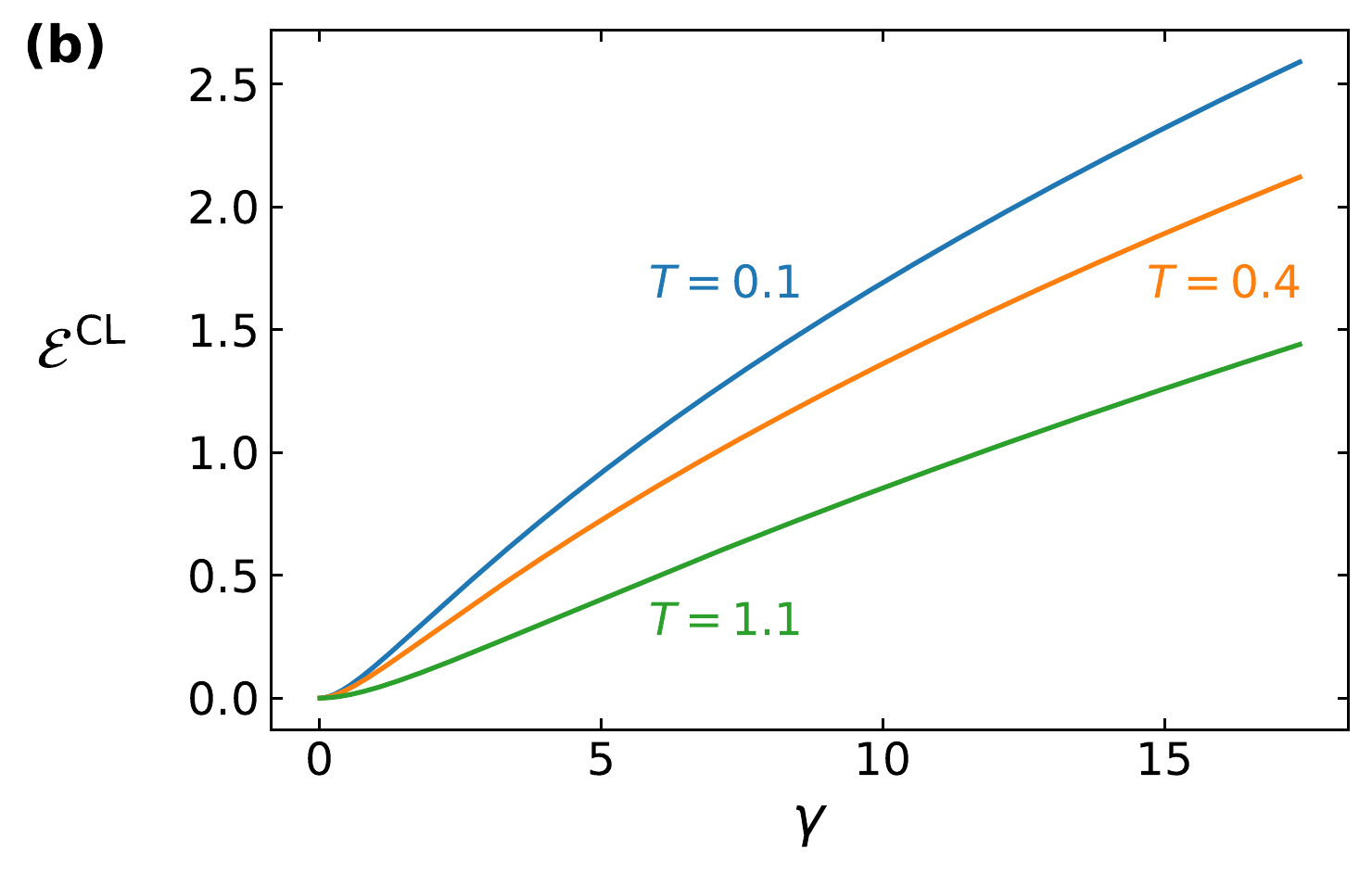}
\caption{Figures of merit of the device as functions of the coupling strength. \textbf{(a)} Efficiency, $\eta^\CL$, and \textbf{(b)} ergotropy, $\mathcal{E}^\CL$, as functions of $\gamma$, for different values of the temperature $T$. The cutoff function is of Lorent-Drude form and $\omega_0 = 2$, $\omega_c = 4$.}
\label{fig1}
\end{figure}

In the weak-coupling limit, namely, when $\gamma \ll 1$, in Appendix~\ref{app:weaksigma}, we prove that, up to the first order in $\gamma$, $\sigma_{11}^\infty = \sigma_{11}^{\mathrm{(free)}} + \frac{\Phi_T}{2 \pi \omega_0} \gamma$ and $\sigma_{22}^\infty = \sigma_{22}^{\mathrm{(free)}} + \frac{\omega_0 \Psi_T}{2 \pi} \gamma$, where $\Phi_T$ and $\Psi_T$ are dimensionless functions of $\omega_0$, $\omega_c$, and $T$ (see Eqs.~\eqref{Phi} and \eqref{Psi}). Using these in Eqs.~\eqref{ergo:simp} and \eqref{wcd:eq}, we find that $W_\cd^\CL \propto \omega_0 \gamma$ and
\bea \label{weakEeta}
\mathcal{E}^\CL \propto \omega_0 \gamma^2, \quad \Longrightarrow \quad \eta^\CL \propto \gamma;
\eea
see Appendix~\ref{app:weakEW} for details (as well as a discussion on the low-temperature limit). The message of Eq.~\eqref{weakEeta} is that the device is essentially useless in the weak coupling limit: not only the output is small---which is indeed expected in this regime---but also the efficiency is vanishing.

Interestingly, the ultrastrong-coupling limit ($\gamma \to \infty$) is not much better: the efficiency also goes to zero, this time, $\propto \gamma^{-1/2}$. Indeed, in Appendix~\ref{app:strongsigma} we show that, when $\gamma \gg \omega_c^2 / \omega_0^2$, $\sigma^\infty_{11} \propto \frac{1}{\omega_0} \gamma^{-1/2}$ and $\sigma^\infty_{22} \propto \omega_0 \gamma^{1/2}$, which, plugged into Eqs.~\eqref{ergo:simp} and \eqref{wcd:eq}, yield $W_\cd^\CL \propto \omega_c \gamma$ and
\bea \label{kale}
\mathcal{E}^\CL \propto \omega_0 \gamma^{1/2}, \quad \Longrightarrow \quad \eta^\CL \propto \gamma^{-1 / 2}.
\eea

With these observations, we expect that the efficiency, as a function of $\gamma$, is maximized at some intermediate value of $\gamma$, which is in fact what we observe in Fig.~\ref{fig1}(a), where, for the Lorentz-Drude cutoff function (Eq.~\eqref{fLor}), the numerically calculated $\eta^\CL$ is plotted against $\gamma$, for three different values of the temperature, $T$. Fig.~\ref{fig1}(b) shows the dependence of $\mathcal{E}^\CL$ on $\gamma$ for the same choice of parameters, and it can be seen how $\mathcal{E}^\CL$ changes its convex ($\gamma^2$) behavior, at small $\gamma$'s, to concave behavior ($\sqrt{\gamma}$), at large $\gamma$'s. Plausibly assuming that the thermalization (hence, charging) time, $t_c$, monotonically decreases with $\gamma$ (see the corresponding discussion in Sec.~\ref{sec:engine}) for all values of $\gamma$, and viewing the device as a thermal machine, we notice that the power per cycle, $\mathcal{E}^\CL/ t_c$, increases with $\gamma$ (see Eq.~\eqref{kale}). Therefore, the assumption holding, there is a certain power-efficiency trade-off: for sufficiently large $\gamma$'s, more power means less efficiency.

An interesting observation from Fig.~\ref{fig1} is that, after its peak, $\eta^\CL$ decays rather slowly with the increase of $\gamma$, which means that the device can maintain a close-to-maximum efficiency while producing a significantly larger amount of output work (ergotropy) than that at maximum efficiency. Indeed, e.g., for $T = 0.1$, the maximum of $\eta^\CL$ ($\approx 6.5 \%$) is reached at $\gamma_{\max} \approx 3.8$, with $\mathcal{E}^\CL(\gamma_{\max}) \approx 0.7$, whereas at $\gamma = 14.4$, the device still operates at $\approx 80 \%$ of the maximal efficiency, while outputting $3.2$ times as much ergotropy compared to that at $\gamma_{\max}$.

We also observe from Fig.~\ref{fig1} that both $\eta^\CL$ and $\mathcal{E}^\CL$ are \textit{decreasing} as the temperature increases. It can also be checked that $W_\cd^\CL$ is an \textit{increasing} function of $T$. In fact, by numerically checking a wide range of parameters and several cutoff functions, we found that these monotonicities are a general feature of the Caldeira-Leggett model. Moreover, as we prove in Appendix~\ref{app:sigmahighT} for an arbitrary Caldeaira-Leggett model, at high temperatures ($T \gg \omega_c$), energy equipartition holds: $\omega_0^2 \sigma_{11}^\infty = T + O \Big(\frac{\omega^2_0}{T} \Big)$ and $\sigma_{22}^\infty = T + O \Big( \frac{\omega^2_0}{T} \Big)$. Through Eqs.~\eqref{ergo:simp} and \eqref{wcd:eq}, these yield
\bea \nonumber
W_\cd^\CL = \frac{\omega_R^2}{\omega_0^2} T + O \bigg( \frac{\omega^2_0}{T}\bigg) \quad \mathrm{and} \quad \mathcal{E}^\CL = O \bigg( \frac{\omega_0^4}{T^3} \bigg), ~
\eea
and therefore, $\eta^\CL = O \Big( \frac{\omega_0^4}{T^4} \Big)$. We see that the device's figures of merit, $\mathcal{E}^\CL$ and $\eta^\CL$, decay with $T$ very quickly, which means that low temperatures are essential for the efficient operation of the device.

\begin{figure}[t!]
\center
\includegraphics[width = 8.5 cm]{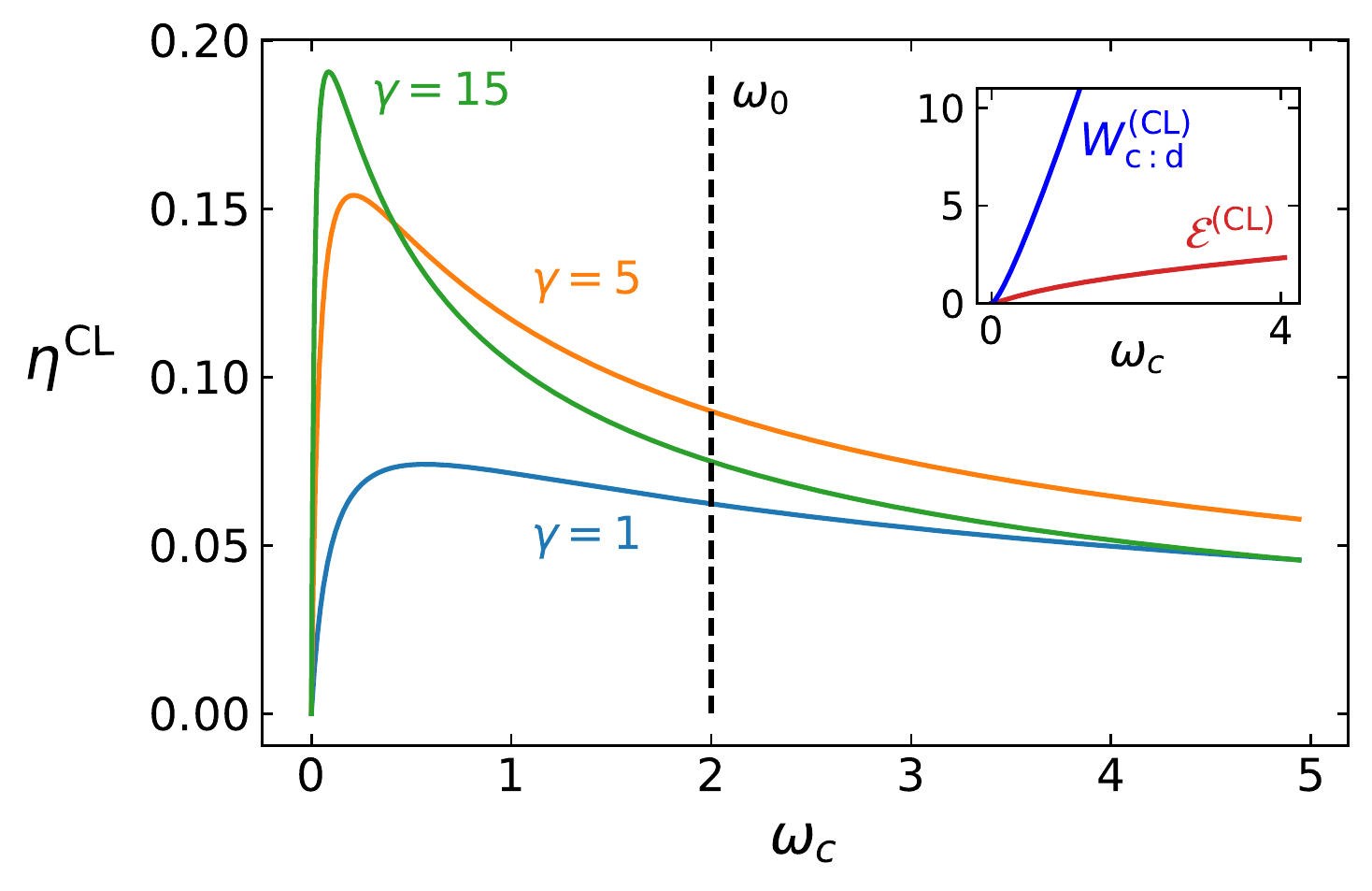}
\caption{Figures of merit of the device as functions of the cutoff frequency. On the \textbf{main plot}, efficiency, $\eta^\CL$ is shown as a function of $\omega_c$, for different values of the coupling constant, $\gamma$. Observe that the higher the desired efficiency, the sharper the peak and the smaller the $\omega_c$ have to be. The \textbf{inset}, where $\mathcal{E}^\CL$ and $W^\CL_\cd$ are plotted against $\omega_c$, for $\gamma = 15$, is to illustrate that, in contrast to the efficiency, $\mathcal{E}^\CL$ and $W^\CL_\cd$ are monotonically increasing, respectively, concave and convex functions of $\omega_c$. The cutoff function is of Lorentz-Drude form and $\omega_0 = 2$, $T = 0.1$.}
\label{fig2}
\end{figure}

We can see in Fig.~\ref{fig1} that the device operates with a rather small efficiency. Looking to find high-efficiency regimes, we turn to $\omega_c$---the only parameter we have not explored yet. Generically, it is assumed that $\omega_c$ is large enough to be greater than $\omega_0$, and it is the choice made in Fig.~\ref{fig1}: $\omega_0 = 2$ and $\omega_c = 4$. However, it turns out that the maximal---over all the parameters---efficiency the Caldeira-Legget model can allow, with the condition that $\omega_c \geq \omega_0$, is $\approx 10.19 \%$, which is achieved at $T = 0$, $\gamma \approx 5.94$, and $\omega_c = \omega_0$, and does not depend on the value of $\omega_0$, as long as it is $> 0$. Indeed, we already knew that $\eta^\CL$ is a monotonically decreasing function of $T$, so $T = 0$ follows trivially. With $T = 0$, the only dimensional parameters are $\omega_0$ and $\omega_c$, therefore, the dimensionless $\eta^\CL$ can depend only on $\tom_0 = \omega_0 / \omega_c$. We find numerically that, for sufficiently large values of $\gamma$, $\eta^\CL$ is an increasing function of $\tom_0$, for $\tom_0 \leq 1$; this fact is illustrated in Fig.~\ref{fig2}. The only free parameter left now is $\gamma$, and we find the above-mentioned optimal values of $\gamma$ and $\eta^\CL$ also numerically. To sum up, if we want efficiencies higher than $10 \%$, we need to consider $\omega_c < \omega_0$. Such situations may occur when the bath is a harmonic (Rubin) chain with nearest-neighbour interactions \cite{Hovhannisyan_2018}, and the individual frequencies of the oscillators in the chain are smaller than $\omega_0$. In Fig.~\ref{fig2}, it is shown that, for sufficiently high $\gamma$'s, the efficiency does go above the $10 \%$ value. The nonmonotonic behavior of $\eta^\CL$ is due to the fact that $\mathcal{E}^\CL$ is concave whereas $W^\CL_\cd$ is convex, although both $\mathcal{E}^\CL$ are $W^\CL_\cd$ monotonic in $\omega_c$ (see the inset of Fig.~\ref{fig2}). Importantly, high efficiencies are obtained only at small values of $\omega_c$, which means that the output work also has to be small. Moreover, with the increase of $\gamma$, the peak around the maximum becomes increasingly sharper, which means that the cutoff frequency of the bath needs to be fine-tuned in order to achieve higher efficiencies. The efficiency can be further increased by taking larger $\omega_0$'s and $\gamma$'s; we numerically found that the highest efficiency possible within this model is $50 \%$, which is achieved in the $\tom_0 \to \infty$ and $\gamma \to \infty$ limit.

\subsection{Several identical oscillators attached to a common bath}
\label{sec:manycopy}

As we saw in the preceding subsection, in order for the device to produce a significant output ergotropy with a reasonably high efficiency, large values of the coupling constant are necessary, which may be challenging to achieve in practice. Here, we discuss a collective enhancement effect appearing when the system is comprised by $n$ copies of the oscillator, which are simultaneously coupled to a common bath. Below, we will show that this system is essentially equivalent to a single oscillator coupled to the bath, but with a rescaled coupling constant: $n \gamma$.

Indeed, the $n$-oscillator Caldeira-Leggett Hamiltonian with a shared bath consists of the system Hamiltonian, $\sum_{a = 1}^n \big[\frac{p_a^2}{2} + \frac{\omega_0^2 q_a^2}{2}\big]$, and the bath Hamiltonian, $H_B^\CL$, comprising together the bare Hamiltonian, and the interaction term: 
\beaa \nonumber
H^{(n\text{-CL})}_I = \frac{1}{2} \omega_R^2 \bigg( \sum_{a = 1}^n q_a \bigg)^2 - \sum_{a = 1}^n q_a \! \sum_k g_k q_k.
\eeaa
By introducing
\bea \label{XandQ1}
x = (q_1, ... , q_n, p_1, ..., p_n), \quad \; Q_1 = \frac{\sum_{a = 1}^n q_a}{\sqrt{n}}, ~~~~
\eea
we can rewrite the total Hamiltonian as
\bea \nonumber
H^{(n\text{-CL})}_\gT \! = \frac{1}{2} x^{\mathrm{T}} M x + \frac{\Omega_R^2 Q_1^2}{2} - Q_1 \sum_k G_k q_k + H^\CL_B, ~~
\eea
where
\bea \label{h}
M = \left( \omega_0^2 \, \id_n \right) \oplus \id_n, \quad \Omega_R^2 = n \omega_R^2, \quad G_k = g_k \sqrt{n}, ~~~~~
\eea
with $\id_n$ being the $n \times n$ identity matrix and the symbol $\oplus$ denoting the direct sum. Note that the indices $a$ and $b$ will, in this subsection, distinguish the system degrees of freedom from those of the bath, labeled by $k$.

Now, we introduce a change in the system's variables, according to $Q_a = \sum_b \aleph_{ab} q_b$ and $P_a = \sum_b \aleph_{ab} p_b$, with $\aleph$ being an orthogonal matrix ($\aleph \aleph^\mathrm{T} = \id_n$) such that $Q_1 = \sum_a \aleph_{1a} q_a$ coincides with that defined in Eq.~\eqref{XandQ1}. (Note that this condition fixes \textit{only} the first row of $\aleph$.) Given the special structure of $M$ (Eq.~\eqref{h}), this transformation leaves $M$ unchanged. Therefore, in terms of the new system variables, the total system-bath Hamiltonian will be
\beaa \label{nCL3}
H^{(n\text{-CL})}_\gT =& \sum_{a = 1}^{n - 1} \bigg[ \frac{1}{2} P_a^2 + \frac{1}{2} \omega_0^2 Q_a^2 \bigg] + \frac{1}{2} P_1^2 + \frac{1}{2} \omega_0^2 Q_1^2 ~~~~
\\
&+ \frac{1}{2} \Omega_R^2 Q_1^2 - Q_1 \sum_k G_k q_k + H^\CL_B,
\eeaa
which means that the original total system is equivalent to a collection of $n - 1$ free oscillators and a single oscillator coupled to a bath.

As per the initial state of the system, given the symmetry of the problem, we choose it to be a product of identical single-oscillator states, and the initial system-bath state is, as usual, given by Eq.~\eqref{inista}. Moreover, we will assume the single-oscillator states to be purely quadratic Gaussian, so that
\bea \nonumber
\Omega^{(n\text{-CL})} = \bigotimes_{a = 1}^n \gimel(A, B, C, q_a, p_a) \otimes \tau_B^\CL,
\eea
with
\bea \label{gimel}
\gimel(A, B, C, q_a, p_a) \propto e^{- A q_a^2 - B p_a^2 - \frac{1}{2} C \{ q_a, p_a \}},
\eea
where $A$, $B$, and $C$ are real, subject only to the condition that the operator $A q_a^2 + B p_a^2 + \frac{1}{2} C \{ q_a, p_a \} > 0$. Noting that, due to the orthogonality of $\aleph$, $\sum_a q_a^2 = \sum_a Q_a^2$, $\sum_a p_a^2 = \sum_a P_a^2$, and $\sum_a \{ q_a, p_a \} = \sum_a \{ Q_a, P_a \}$, the initial state looks the same from the perspective of the new variables:
\bea \label{grmada}
\Omega^{(n\text{-CL})} = \bigotimes_{a = 1}^n \gimel(A, B, C, Q_a, P_a) \otimes \tau_B^\CL.
\eea
Let us note at this point that, although all states of the form \eqref{gimel} have identical spectra, they all live in different subspaces of the system's Hilbert space.

Summing up Eqs.~\eqref{nCL3} and \eqref{grmada}, the model consists of $n$ identical oscillators with frequency $\omega_0$, $n - 1$ (numbers $2$ to $n$) of which are uncorrelated and uncoupled from the first one and from the bath, whereas the first one evolves as a standard quantum Brownian particle with rescaled coupling: $g_k \to g_k \sqrt{n}$. The states of the uncoupled oscillators evolve under the influence of their internal Hamiltonian (and therefore (i) they do not relax and (ii) their ergotropy is constant in time) and remain uncorrelated to the state of the first particle (which thermalizes with the bath).

Moreover, as we show in Appendix~\ref{app:GErgotropy}, the maximal work cyclic (in Hamiltonian) Gaussian operations can extract from a (multimode) Gaussian state---the ``Gaussian ergotropy''---is given by
\bea \label{repakan}
\mathcal{G} = \frac{1}{2} \tr(\sigma M) - \sum_a s_a^\uparrow m_a^\downarrow,
\eea
where $\{ s_a^\uparrow \}$ are the symplectic eigenvalues of the covariance matrix, $\sigma$ (defined in the multimode case identically to the single-mode case given in Eq.~\eqref{lobzik}), taken in increasing order, and $\{ m_a^\downarrow \}$ are those of $M$, taken in decreasing order (see Appendix~\ref{app:GErgotropy} for detailed definitions and a proof of Eq.~\eqref{repakan}). Now, since all $n$ symplectic eigenvalues of $M$ are equal to each other (all are $\omega_0$), the ergotropy of the $n$-oscillator system is equal to the sum of individual ergotropies. Therefore, our protocol (Eqs.~\eqref{step1}--\eqref{step1p}) will independently process the first (Brownian) particle and the rest of the $n - 1$ particles. Thus, the latter will have no role in the energetics at all, because we will extract all the erogtropy from them during the first cycle---which will be nonzero only if $\gimel(A, B, C, Q_a, P_a)$ is active with respect to $\frac{1}{2} P_a^2 + \frac{1}{2} \omega_0^2 Q_a^2$---and for the rest of the time, they will just remain in their passive states. The energetics for the Brownian particle ($Q_1$, $P_1$), on the other hand, will be identically as described in Sec.~\ref{sec:CLenergetics}, with the only difference that, since $G_k = g_k \sqrt{n}$ and $J(\omega)$ is quadratic in $g_k$ (Eq.~\eqref{jorik}), the coupling constant is now $n \gamma$.

\section{Discussion}

In this paper, we investigated the arguably simplest model of charging a battery: a system in strong contact with a bath, jointly evolving towards global thermal equilibrium (or a nonthermal state appearing to be thermal when viewed locally). The idea is based on the basic observation that the reduced state of a subsystem of a globally thermal system is not generally thermal and may thus harbor extractable work when disconnected from the global system. This setup presents a number of benefits, not simultaneously met in any other battery-charger setup, such as no need of fine control over the preparation of the charged state and no necessity to exert any effort to maintain the battery in that state. Indeed, the existing setups either require isolating the system or engineering a special system-bath evolution \cite{Campaioli_2018, Santos_2019, Barra_2019} or maintaining fragile internal symmetries \cite{Hovhannisyan_2019, Liu_2019} or active external stabilization in order to preserve the charged state \cite{Gherardini_2020}.

We studied this setup in full generality, revealing fundamental limitations on the process, encapsulated in a nonnegative entropy production (Eq.~\eqref{final1}), quantifying the amount of work that needs to be dissipated in order for the device to function. A key peculiarity of the charging cycle is that, nonweak coupling, being responsible for preparing the charged state of the battery, is also an inevitable source of dissipation. Indeed, at least the disconnecting stroke must be fast, because, otherwise, by the end of the disconnecting stroke, the system will be in thermal equilibrium with the bath while being only weakly coupled to it, i.e., it will be in a Gibbs state, which is passive. Therefore, the device is expected to function optimally in the moderate-to-large coupling regime, which is what we saw on the example of Caldeira-Leggett model in Sec.~\ref{sec:CLenergetics}.

Rephrasing the previous paragraph, the autonomy and robustness of our device come at the cost of limited efficiency. However, it is important to note that $\eta$ is merely an upper bound to the ``\textit{de facto} efficiency'', which we define to be one that compares $\mathcal{E}$ to the \textit{actual energy spent} on the charging cycle, including the energy spent on control and stabilization. To illustrate this point, observe that one can simply take a system in the ground state and call it a depleted battery; then, one can use a unitary transformation to rotate that ground state to the highest energy eigenstate of that system, thereby charging the battery. The ``formal efficiency'' of that charging cycle is $1$---no energy is dissipated and all energy transferred to the battery is unitarily extractable \footnote{The problem of dissipationlessly extracting the work stored in the battery is a separate, nontrivial issue, actively studied in the literature \cite{Woods_2019, Monsel_2020}.}. Immersed into reality, this simple picture breaks down by the necessities of isolating the battery and performing tailored unitary operations on it. All this requires an intervention of macroscopic laboratory equipment that consumes macroscopic amounts of energy. Moreover, keeping an excited system from spontaneously emitting a photon (i.e., leaking the charge) also requires complicated equipment, thus also requiring macroscopic energy expenditures. With all these taken into account, the de facto efficiency of that simple battery will essentially be zero. Although our design is not completely autonomous in that it is sensitive to the bath's temperature and certain, albeit little, external control is required, it is free from two major sources of macroscopic energy expenditure described above. Therefore, the de facto efficiency of our design is arguably higher than that of other existing designs, which are at best free from only one of such sources of energy expenditure and therefore suffer close-to-zero de facto efficiency. Moreover, to the best of our knowledge, except for this work, the thermodynamic efficiency of charging and discharging process has been addressed only in Ref.~\cite{Barra_2019}, albeit in a slightly different setting.

In our cycle, the connection and disconnection steps are instantaneous, therefore, the minimum time needed to run the cycle is determined by the relaxation time, which in turn depends on the strength of the coupling. As we briefly mentioned in Sec.~\ref{sec:discE}, other time-dependent protocols for connecting and disconnecting the system could be considered, exploring whether (and ultimately to what extent) the efficiency can increase and whether that occurs at the expense of the time it takes to run the cycle.

Another aspect of the device is that the ``battery'' system has to be small. Indeed, the athermality of a macroscopic system coupled to a bath will generically be a boundary effect tending to thermalize away once the system is detached from the bath. Moreover, generically, the effect of system-bath correlations due to interaction are more pronounced at low temperatures. Therefore, although the working principle of our device is not inherently quantum, it is better suited for the quantum regime. This aspect is well illustrated by the Caldeira-Leggett model, for which the machine is most efficient in the low-temperature regime, where its quantum features are most prominent. In the classical (high temperature) regime, the system exhibits energy equipartition, independently from the details of the bath and the strength of the coupling, so that the ergotropy vanishes. We note however that, in this work, we forewent studying quantum effects such as system-bath entanglement and coherence in the system's state, leaving the study of these important questions to the future.

In a simplistic scenario of $n$ noninteracting oscillators coupled to a common bath, in Sec.~\ref{sec:manycopy}, we showed that enhancement of both the output and efficiency takes place in the Caldeira-Leggett model. Situations where collective effects are more pronounced, e.g., when the ``working medium'' is critical \cite{Campisi_2016, Sune_2019prl}, constitute a promising direction of future research.

Lastly, we envision that our battery-charger setup can be experimentally realized on certain quantum optical platforms where strong and ultrastrong coupling regimes are reachable \cite{Kockum_2019, Felicetti_2020}, such as, e.g., cavity QED \cite{Niemczyk_2010, Forn-Diaz_2017}. Another possible direction to look for practical realizations of our protocol, is using chemical bonds, which are a good example of naturally occurring strongly coupled microscopic systems. For example, the cycle can be realized through creating chemical bonds, which can store the energy for a long time, and then breaking them and using the athermal states of the components to extract work.

\section*{{ACKNOWLEDGMENTS}}

We thank Jerzy \L{}uczka for making us aware of Ref.~\cite{Bialas_2018} after the first preprint version of this work has appeared on arXiv.org. F. B. thanks Fondecyt project 1191441 and the Millennium Nucleus ``Physics of active matter'' of the Millennium Scientific Initiative. A. I. was supported by the Danish Council for Independent Research and the Villum Foundation. A. I. gratefully acknowledges the hospitality of the Department of Physics at University of Chile and the financial support from Fondecyt project 1191441 during his stay at University of Chile, where part of this project was initiated.

\appendix


\section{{THE DEFINITION OF ERGOTROPY}}
\label{app:ergdef}

In this appendix, we briefly outline well-known results about passivity \cite{Pusz_1978, Lenard_1978} and ergotropy \cite{Allahverdyan_2004}.

Ergotropy is defined as the maximal amount of work extractable from a system by means of a cyclic Hamiltonian process \cite{Allahverdyan_2004}. Namely, given a system in a $d$-dimensional Hilbert space in some state $\rho$ and Hamiltonian $H = \sum_k E_k \ket{k}\bra{k}$ at the initial moment of time $t = t_{in}$, one drives the system Hamiltonian according to some time-dependent protocol $H(t)$ in such a way that, at the end of the process (the final moment of time $t = t_{fin}$) the system Hamiltonian is back to its original value: $H(t_{in}) = H(t_{fin}) = H$. Such processes are called cyclic Hamiltonian processes, and their fundamental importance in thermodynamics is that, at the end of the process, we end up with the \textit{same system} as we had at the beginning of the process. The Kelvin-Planck formulation of the Second Law refers to these very processes. Indeed, if one considers processes at the end of which the Hamiltonian is allowed to differ from the initial Hamiltonian, then extracting work from a thermal system becomes trivial: imagine an initially thermal gas in a chamber expanding adiabatically and pushing against a piston---this process extracts work from a system in thermal equilibrium, as long as one does not require the piston to be back at its original position at the end of the process.

The time evolution of the state under the influence of the time-dependent Hamiltonian is unitary: for $\forall t \in [t_{in}, t_{fin}]$,
\bea
\rho(t) = U(t, t_{in}) \rho U(t, t_{in})^\dagger,
\eea
where the unitary evolution operator is standardly written as
\bea
U(t, t_{in}) = \mathcal{T} e^{-\ci \int_{t_{in}}^t ds H(s)},
\eea
the symbol $\mathcal{T}$ signifying chronological ordering. The inverse is also true: any unitary evolution can be generated by a time-dependent Hamiltonian process. Indeed, given a unitary $U(t_{fin}, t_{in})$, we can generate it by the cyclic Hamiltonian process where we, at the moment of time $t_{in}$ abruptly change the Hamiltonian from $H$ to $\frac{\ci}{t_{fin} - t_{in}}\ln U(t_{fin}, t_{in})$, let it run until the moment of time $t_{fin}$, and then abruptly change the Hamiltonian back to $H$. In view of this, the ergotropy of a system in state $\rho$ with respect to Hamiltonian $H$ can be defined as
\bea \label{erg1}
\mathcal{E} = \mathcal{E} (\rho, H) = \tr[H \rho] - \min_{U} \tr[H U \rho U^\dagger],
\eea
therefore, $\mathcal{E}$ is sometimes also called unitarily extractable work.

The unitary evolution operator delivering the minimum in Eq.~\eqref{erg1},
\bea \label{erg2}
U_{\mathcal{E}} := \arg\min_{U} \tr[H U \rho U^\dagger],
\eea
thus takes the system to a state from which no more work can be extracted (by means of cyclic Hamiltonian processes). The latter, $\rho_p = U_{\mathcal{E}} \rho U_{\mathcal{E}}^\dagger$, is called \textit{passive} \cite{Pusz_1978, Lenard_1978}, and it can be shown that, in the eigenbasis of $H$ with the basis elements chosen such that the eigenvalues of $H$ are in the increasing order: $E_1 \leq E_2 \cdots \leq E_d$,
\bea \label{erg3}
\rho_p = \mathrm{diag} \p{r_1^\downarrow, r_2^\downarrow, \cdots, r_d^\downarrow},
\eea
where $r_k^\downarrow$ are the eigenvalues of $\rho$ in the decreasing order ($r_1^\downarrow \geq \cdots \geq r_d^\downarrow$). Obviously, since $\rho_p$ is diagonal in the eigenbasis of $H$, $[\rho_p, H] = 0$. Also note that, since all positive-temperature Gibbs states over $H$ are already in the form \eqref{erg3}, they are all passive; however, not all passive states are Gibbs states (see the discussion about complete passivity in Refs.~\cite{Pusz_1978, Lenard_1978}). Lastly, let us note that the ergotropy of a system is invariant under the internal dynamics. Indeed, $\tr \big[ H e^{- \ci H t} \rho e^{\ci H t} \big] = \tr[H \rho]$ and, since $e^{- \ci H t}$ is unitary, $\min\limits_{U} \tr \big[H U e^{- \ci H t} \rho e^{\ci H t} U^\dagger \big] = \min\limits_{U} \tr[H U \rho U^\dagger]$, so $\mathcal{E} \p{e^{- \ci H t} \rho e^{\ci H t}, H} = \mathcal{E}(\rho, H)$.

\section{{THERMALIZATION IN NONCRITICAL MANY-BODY SYSTEMS WITH SHORT-RANGE INTERACTIONS}}
\label{app:thermalization}

Generally, the results on thermalization are obtained in two steps. One first proves that the system equilibrates, i.e.,
\bea \nonumber
\nu_\mathfrak{S} (\av{\Omega}) := \lim\limits_{\mathcal{T} \to \infty} \frac{1}{\mathcal{T}} \int_0^{\mathcal{T}} \! dt' \Vert \tr_{\mathfrak{T} \backslash \mathfrak{S}} \Omega_{t'} - \tr_{\mathfrak{T} \backslash \mathfrak{S}} \av{\Omega} \Vert, ~
\eea
is small. Here, $\av{\Omega} = \lim\limits_{\mathcal{T} \to \infty} \frac{1}{\mathcal{T}} \int_0^{\mathcal{T}} dt \Omega_t$. In order to establish that, one uses the general bounds on equilibration obtained in Refs.~\cite{Linden_2009, Short_2012, Garcia-Pintos_2017}; most appropriately in our situation \cite{Short_2012},
\bea \label{equilib}
\nu_\mathfrak{S} (\av{\Omega}) \leq \sqrt{\frac{d_\mathfrak{S}^2 G_D}{d_\mathrm{eff}}},
\eea
where $d_\mathfrak{S}$ is the Hilbert-space dimension of the subsystem $\mathfrak{S}$, $G_D$ is the highest ``gap degeneracy'' of $H_\gT$ (gap degeneracy is the number of times a given gap is repeated in the spectrum, and $G_D$ is the largest such number), and $d_\mathrm{eff}$, the ``effective dimension'', is
\bea
d_\mathrm{eff} = \frac{1}{\sum_a \big[ \tr(\Omega P_a) \big]^2},
\eea
where $P_a$ are the eigenprojectors of $H_\gT$ ($H_\gT = \sum_a E_a P_a$, where $E_a$ are $H_\gT$'s eigenvalues). When, as in our setting, $d_\mathfrak{S}$ is a fixed finite number, the thermalization is guaranteed as long as $\frac{G_D}{d_\mathrm{eff}}$ is $\ll 1$ and tends to zero with $N \to \infty$. Generically, for canonical and microcanonical $R_B$'s, one expects $d_\mathrm{eff}$ to be exponential in $N$ in view of the exponential (in $N$) density of eigenvalues of $H_\gT$ and $H_B$ \cite{Reimann_2008}. However, that is not always the case; instead, in Ref.~\cite{Farrelly_2017}, it was shown that, whenever $H_\gT$ is $k$-local and $\Omega$ has exponentially decaying correlations, which, since in our case $\Omega = \rho_0 \otimes R_B$, is guaranteed as long as $R_B$ has exponentially decaying correlations,
\bea \label{FBC1}
d_\mathrm{eff} \geq O \bigg( \frac{\sqrt{N}}{\ln^{2 d} N} \bigg),
\eea
where $d$ is the spatial dimension of the lattice, meaning that equilibration is guaranteed as long as $G_D$ does not scale with $N$ (or scales slower than $\sqrt{N}$), which is expected to not hold only in extremely exotic cases as it holds in all known models and, in general, Hamiltonians with degenerate gaps are of zero measure in the space of all Hamiltonians and any given degeneracy can be lifted by an infinitesimal perturbation. See also Ref.~\cite{Garcia-Pintos_2017} for a discussion on equilibration time-scales.

Another important bound proven in Ref.~\cite{Brandao_2015}, and adapted in Ref.~\cite{Farrelly_2017}, is the following. Say, $\tau$ is some state with exponentially decaying correlations on the lattice and $0 < \alpha < 1$. Then, for any state $\rho$ satisfying
\bea \label{BC1}
S(\rho \Vert \tau) = \circ \big( N^{\frac{\alpha}{d + 1}}\big),
\eea
it holds that
\bea \label{BC2}
\mathbbm{E}_m \Vert \tr_{\mathfrak{T} \backslash \mathfrak{S}_m} \rho - \tr_{\mathfrak{T} \backslash \mathfrak{S}_m} \tau \Vert \leq O \big( N^{- \frac{1 - \alpha}{2 d + 4}} \big),
\eea
where $\mathbbm{E}_m$ denotes the arithmetic mean over all subsystems $\mathfrak{S}_m$ of diameter $m$ lattice units. For this bound to be valid, $m$ must be $\circ \Big( N^{\frac{d + 1 + \alpha}{(d + 1) (d + 2)}} \Big)$, which is the case in our analysis as we will always work with subsystems that do not scale with $N$, i.e., $m$ will always be a finite number. Note that the bound \eqref{BC2} holds on average, whereas in this paper we need it to hold only for all subsystems in $s \cup \mathrm{supp}(H_I)$. Importantly, we do not require \eqref{BC2} to be valid for all subsystems of $\gT$, so we do not need to require translation invariance of $\gT$. However, $H_\gT$ must satisfy certain transport properties such as nonzero Lieb-Robinson velocity \cite{Nachtergaele_2010}. Indeed, in systems with inhibited energy/information transport, e.g., those that are many-body localized, there will be subsystems which ``remember'' their initial states (see Ref.~\cite{Gogolin_2016} for a discussion); here, we require that no localization phenomena occur on (or near) $s \cup \mathrm{supp}(H_I)$.

Combining Eqs.~\eqref{equilib}, \eqref{FBC1}, and \eqref{BC2}, and using the triangle inequality for the trace norm \cite{mikeike}, one obtains \cite{Farrelly_2017}
\bea \label{FBC2}
\nu_\mathfrak{S}(\tau_\gT) \leq O \big( G_D^{1/2} N^{-1/4} \ln^d N \big) + O \big( N^{-\frac{1 - \alpha}{2 d + 4}} \big), ~~~
\eea
whenever $S(\av{\Omega} \Vert \tau_\mathfrak{T}) = \circ \big( N^{\frac{\alpha}{d + 1}} \big)$. Since $\av{\Omega}$ is obtained from $\Omega$ by erasing some of its nondiagonal elements in $H_\gT$'s eigenbasis, $S(\av{\Omega}) \geq S(\Omega)$. Also, by $\av{\Omega}$'s definition, it holds that $\tr(H_\gT \av{\Omega}) = \tr(H_\gT \Omega)$. Therefore, $S(\av{\Omega} \Vert \tau_\mathfrak{T}) \leq S(\Omega \Vert \tau_\gT)$, meaning that the bound \eqref{FBC2} holds as long as
\bea \label{FBC3}
S(\Omega \Vert \tau_\gT) = \circ \big( N^{\frac{\alpha}{d + 1}} \big).
\eea

Now, when the initial state of the bath is $\tau_B$, for any $\rho_0$, we can write
\bea
\Omega_c := \rho_0 \otimes \tau_B = \frac{e^{-\beta \overline{H}}}{\tr e^{-\beta \overline{H}}},
\eea
with
\bea
\overline{H} = - T \ln \rho_0 + H_B,
\eea
and hence,
\bea \label{PB1}
S(\Omega_c \Vert \tau_\gT) &=& \beta \tr((H_\gT - \overline{H}) \Omega_c) + \ln \frac{\tr e^{-\beta H_\gT}}{\tr e^{-\beta \overline{H}}}
\\ \nonumber
&\leq& \beta \tr((H_\gT - \overline{H}) \Omega_c) - \beta \tr((H_\gT - \overline{H}) \tau_\gT) ~~~~
\\ \nonumber
&\leq& 2 \beta \Vert H_\gT - \overline{H} \Vert
\\ \label{PB2}
&=& 2 \beta \Vert H_s + T \ln \rho_0 + H_I \Vert = O(1),
\eea
where the first inequality is due to the Peierls-Bogoliubov inequality \cite{Wehrl_1978}, the second inequality is by the very definition of the trace norm \cite{mikeike}, and last equality is due to the fact that $H_I$ is at most $k$-local (because $H_\gT$ is). Hence, Eq.~\eqref{BC2} is satisfied with $\alpha = 0$, which, in view of Eq.~\eqref{BC1}, means that $\nu_\mathfrak{S}(\tau_\gT)$ goes to zero polynomially with $N$. In other words, when the bath starts in a canonical state ($R_B = \tau_B$), thermalization as in Eq.~\eqref{thermaliz} takes place. 

Using the results in Refs.~\cite{Brandao_2015, Tasaki_2018} about the equivalence of canonical and microcanonical ensembles, and combining them with results in Ref.~\cite{Farrelly_2017} mentioned above, we will now show that, with the same requirements on $H_\gT$ as above, thermalization as in Eq.~\eqref{thermaliz} takes place also when the bath starts in a microcanonical state, $\mu(E, \Delta)$, as defined in Eq.~\eqref{microcan}. To ensure that $\tau_B$ and $\mu_B(E, \Delta)$ (with $E = \tr(H_B \tau_B)$) are equivalent, we choose \cite{Brandao_2015}
\bea
O \big( \ln^{2d} N \big) \leq \Delta \leq O \big(\sqrt{N} \big).
\eea
For such a choice, as is proven in Refs.~\cite{Brandao_2015, Tasaki_2018},
\bea \label{indist}
\Vert \tr_{B \backslash \mathfrak{S}} \mu_B(E, \Delta) - \tr_{B \backslash \mathfrak{S}} \tau_B \Vert \leq O \bigg( \frac{\ln^d N}{N^{\frac{1}{2 d + 4}}} \bigg), ~~~~~
\eea
where $\mathfrak{S}$ is an arbitrary subsystem of $B$ with a diameter $\leq O \Big( N^{\frac{1}{d (d + 1)}}\Big)$.

In order to prove that the system equilibrates, we need to ensure that $d_\mathrm{eff}$ is large. To see that it is, we note that, in view of local indistinguishability of $\mu_B(E, \Delta)$ and $\tau_B$ (Eq.~\eqref{indist}), the correlators on $\mu_B(E, \Delta)$ and $\tau_B$ will coincide up to an $O \bigg( \frac{\ln^d N}{N^{\frac{1}{2 d + 4}}} \bigg)$ correction, meaning that, up to distances $\leq O (\ln N)$, correlations in $\mu_B(E, \Delta)$ are guaranteed to decay exponentially, which is sufficient to prove the bound \eqref{FBC1} \cite{Farrelly_2017}. In systems where the density of eigenstates of the bath increases exponentially with $N$, $d_\mathrm{eff}$ is obviously guaranteed to also be large (keeping in mind that $\Vert H_I \Vert$ does not scale with $N$, see also the discussion in Ref.~\cite{Garcia-Pintos_2017}).

Having established equilibration, we now need to bound $S(\Omega_{mc} \Vert \tau_\gT)$, where $\Omega_{mc} = \rho_0 \otimes \mu_B(E, \Delta)$:
\bea \nonumber
S(\Omega_{mc} \Vert \tau_\gT) &=& \tr(\Omega_{mc} \ln \Omega_{mc}) - \tr(\Omega_{mc} \ln \tau_\gT)
\\ \nonumber
&& \qquad + \left[ \tr(\Omega_{mc} \ln \Omega_c) - \tr(\Omega_{mc} \ln \Omega_c) \right]
\\ \nonumber
&=& S(\mu_B(E, \Delta) \Vert \tau_B) + \beta \tr(\Omega_{mc} (H_\gT - \overline{H}))
\\ \nonumber
&& \qquad + \ln \frac{\tr e^{-\beta H_\gT}}{\tr e^{-\beta \overline{H}}}.
\eea
Again, using the Peierls-Bogoliubov inequality in the last term and the definition of trace norm as we did in Eqs.~\eqref{PB1}--\eqref{PB2}, we get that
\bea \nonumber
S(\Omega_{mc} \Vert \tau_\gT) \leq S(\mu_B(E, \Delta) \Vert \tau_B) + 2 \beta \Vert H_s + T \ln \rho_0 + H_I \Vert.
\eea
Finally, invoking the following inequality from Ref.~\cite{Brandao_2015} (Lemma 7):
\bea
S(\mu_B(E, \Delta) \Vert \tau_B) \leq O \big( \ln^{2d} N \big),
\eea
we obtain that $S(\Omega_{mc} \Vert \tau_\gT) \leq O \big( \ln^{2d} N \big)$, and therefore it satisfies Eq.~\eqref{BC1}. In turn, this means that Eq.~\eqref{FBC2} again holds, implying that thermalization takes place also when the bath starts in a microcanonical state.

\subsubsection*{The case when \texorpdfstring{$\mathrm{supp} (H_I)$}{supchik} is not finite}
\label{app:ThermInfSupp}

We have just shown that, for any $H_I$, as long as $\mathrm{supp} (H_I)$ is \textit{finite}, Eq.~\eqref{thermaliz} generally holds up to a correction $O\left( N^{-\epsilon} \right)$, with some $\epsilon > 0$ which is to be read from Eq.~\eqref{FBC2} (although it is possible to be more specific, it is sufficient to note that, when $G_D$ scales at most as $N^y$, with some $0 \leq y < 1/2$, $O\p{N^{- \epsilon}}$, with an arbitrary $0 < \epsilon < \min\left\{ \frac{1}{4} - \frac{y}{2}, \frac{1 - \alpha}{2 d + 4} \right\}$, will be an upper bound for the corrections to Eq.~\eqref{thermaliz}).

Let us now turn to the case when $\mathrm{supp} (H_I)$ is not finite, e.g., when it scales with $N$. In this case, the $O\left( N^{-\epsilon} \right)$ corrections may potentially sum up into something nonnegligible. Therefore, we consider only such $H_I$'s that can be decomposed into a sum of local terms: $H_I = \sum_\kappa \gamma_\kappa V_\kappa$, where $\gamma_\kappa \geq 0$ are the ``coupling constants'' and, for all $\kappa$'s, the diameter of $\mathrm{supp} (V_\kappa)$ is $< C$, with $C$ being some finite number not depending on $\kappa$. Moreover, $\gamma_\kappa$ are adjusted so that, for any $\kappa$, $\max \{ |\tr(V_\kappa \Omega )|, |\tr(V_\kappa \tau_\gT )|\} \leq K$, where $K > 0$ is a finite constant independent on $\kappa$. Then, for each term, the substitution of $R_B$ by $\tau_B$ and $\Omega_\infty$ by $\tau_\gT$ in Eqs.~\eqref{ene1} and \eqref{ene2} introduces an $f(\gamma_\kappa) O\left( N^{-\epsilon} \right)$ correction (here $f$ is some function; see next). Therefore, in order to guarantee that Eq.~\eqref{ene1th} is correct, we require the series $\sum_\kappa f(\gamma_\kappa)$ to scale slower than $N^\epsilon$. Note that this case formalizes the situation when the system is attached to a bosonic bath, as in the Caldeira-Leggett and spin-boson models, where the accumulation of corrections is forestalled by the introduction of a cutoff frequency (see Sec.~\ref{sec:CLmodel} for definitions). There, $V_\kappa$'s correspond to $q \otimes q_k$'s and $\gamma_\kappa$'s to $g_k$'s. Due to the fact that $\tr(q_k \tau_B) = 0$, any nonzero quantity in $\tr(H_I \tau_\gT)$ will be due to system-bath correlations in $\tau_\gT$ stemming from $g_k \neq 0$ \footnote{Except for the system renormalization term, which however is a fixed quantity and therefore does not alter the scaling of the total correction.}, and, given that $H_I$ is itself $\propto g_k$, any such quantity will be $\propto g_k^2$, meaning that $f(\gamma_\kappa) \to g_k^2$. In turn, $\sum_k g_k^2 \approx \int_0^\infty d\omega \omega J(\omega)$, which, for the exponential cutoff, is finite and, for the Lorentz-Drude cutoff, is $\propto \ln N$, which is slow enough to guarantee the correctness of Eq.~\eqref{ene1th}.

\section{{THE DERIVATION OF EQ.~\eqref{final1}}}
\label{app:EPderivation}

Here, we show to get from Eq.~\eqref{promezh2} to Eq.~\eqref{final1}. To do so, we start with Eq.~\eqref{promezh2},
\beaa
T \Sigma = & \, \tr[H_\gT \rho_p \otimes \tau_B] - \tr[H_\gT \tau_\gT]
\\
& \, - \tr[H_B \tau_B] + \tr[H_B \rho_B^\infty],
\eeaa
and adding and subtracting $\tr[H_\gT \tau_\gT']$ to it, we rewrite it as
\beaa \label{promezh5}
T \Sigma = & \, \tr[H_\gT \rho_p \otimes \tau_B] - \tr[H_\gT \tau_\gT'] + \tr[H_\gT \tau_\gT']
\\
& \, - \tr[H_\gT \tau_\gT] - \tr[H_B \tau_B] + \tr[H_B \rho_B^\infty],
\eeaa
where, using that $H_\gT = - T' \ln \tau_\gT' - T' \ln Z_\gT'$, we can transform first two terms in the RHS into $T' \tr[\tau_\gT' \ln \tau_\gT'] - T' \tr[\rho_p \otimes \tau_B \ln \tau_\gT']$. Noticing, in view of Eq.~\eqref{promezh4}, that $\tr[\tau_\gT' \ln \tau_\gT'] = \tr[\rho_p \otimes \tau_B \ln (\rho_p \otimes \tau_B)]$, we arrive at
\beaa \label{promezh6}
T \Sigma = & \, T' S(\rho_p \otimes \tau_B \Vert \tau_\gT') + \tr[H_\gT \tau_\gT']
\\
& \, - \tr[H_\gT \tau_\gT] - \tr[H_B \tau_B] + \tr[H_B \rho_B^\infty],
\eeaa
where $S(\rho || \tau) = \tr [\rho (\ln \rho - \ln \tau)]$ is the relative entropy \cite{mikeike}. Using the identities $H_\gT = -T \ln \tau_\gT - T \ln Z_\gT$ and $H_B = - T \ln \tau_B - T \ln Z_B$, we can further transform Eq.~\eqref{promezh6} into
\beaa \label{promezh7}
T \Sigma = & \, T' S(\rho_p \otimes \tau_B \Vert \tau_\gT') - T \tr[\tau_\gT' \ln \tau_\gT]
\\
& \, - T S(\tau_\gT) - T S(\tau_B) - T \tr[ \rho_B^\infty \ln \tau_B],
\eeaa
were, adding and subtracting $T \tr[\tau_\gT' \ln \tau_\gT']$, we obtain
\beaa \label{promezh8}
T \Sigma = & \, T' S(\rho_p \otimes \tau_B \Vert \tau_\gT') + T S (\tau_\gT' \Vert \tau_\gT) - T S(\tau_\gT) ~~~~
\\
& \, + T S(\tau_\gT') - T S(\tau_B) - T \tr[ \rho_B^\infty \ln \tau_B],
\eeaa
which, by noting Eq.~\eqref{promezh4} and adding and subtracting $T \tr[\rho_B^\infty \ln \rho_B^\infty]$, we finally bring to
\beaa
\Sigma = & \, \frac{T'}{T} S(\rho_p \otimes \tau_B \Vert \tau_\gT') + S (\tau_\gT' \Vert \tau_\gT)
\\
& \, + I_{\tau_\gT}(s : B) + S(\rho_B^\infty \Vert \tau_B),
\eeaa
which is exactly the desired Eq.~\eqref{final1}.

\section{{STEADY STATE OF THE HARMONIC CALDEIRA-LEGGETT MODEL}}
\label{app:textbook}

The harmonic Caldeira-Leggett is described by the Hamiltonian given in Eq.~\eqref{CLham}, and the Heisenberg equations of motion are given by Eq.~\eqref{Heq}. The solution of the set of equations for $q_k$'s in Eq.~\eqref{Heq} can be written as
\beaa \nonumber
q_k (t) = q_k (t_0) \cos \omega_k (t - t_0) + \frac{\dot{q}_k(t_0)}{\omega_k} \sin \omega_k (t - t_0) ~~~~
\\
+ \frac{g_k}{m_k \omega_k} \int_{t_0}^t d s q(s) \sin \omega_k (t - s), ~
\eeaa
which, substituted into the first equation in Eq.~\eqref{Heq}, yields
\bea \label{meq}
\ddot{q} (t) + (\omega_0^2 + \omega_R^2) q (t) - \! \int_{t_0}^\infty \!\! ds q (s) \eta(t - s) = \xi (t), ~~~~~
\eea
where
\bea
\eta(t) &=& \theta(t) \sum_k \frac{g_k^2}{m_k \omega_k} \sin \omega_k t,
\\ \nonumber
\xi(t) &=& \sum_k g_k \! \left[ q_k \cos \omega_k (t - t_0) + \frac{p_k}{m_k \omega_k} \sin \omega_k (t - t_0) \right] \! .
\\ \label{zhesht}
\eea
Here, noting that $t_0$ is the point in time when the description starts, and so it is the point in time at which the Heisenberg-picture operators are initialized by their Shchr\"{o}dinger-picture originals, we substituted $q_k(t_0) = q_k$ and $p_k(t_0) = p_k$. Moreover, we also note that, since the choice of $t_0$ is arbitrary and we are primarily interested in the long-time behavior of the system, for convenience, while always keeping $t_0 \ll - |t|$ finite, we will take the limit $t_0 \to - \infty$ at relevant places.

Using the definition of the spectral density, Eq.~\eqref{jorik}, we can write:
\bea \label{etta}
\eta(t) &=& \frac{2}{\pi} \theta(t) \int_0^\infty d\omega J(\omega) \sin \omega t,
\\ \label{omerInt}
\omega_R^2 &=& \frac{2}{\pi} \int_0^\infty d \omega \frac{J(\omega)}{\omega}.
\eea

Switching to the Fourier-transformed functions ($\overline{q}(\omega) = \int_{-\infty}^\infty dt e^{\ci \omega t} q(t)$) in Eq.~\eqref{meq}, and keeping in mind that $t_0 \to -\infty$, we obtain
\bea
\overline{q}(\omega) = \frac{\overline{\xi}(\omega)}{\alpha(\omega)},
\eea
and hence
\bea
\overline{p}(\omega) = - \ci \omega \frac{\overline{\xi}(\omega)}{\alpha(\omega)},
\eea
where
\bea \label{alfombra}
\alpha(\omega) = \omega_0^2 - \omega^2 + \omega_R^2 - \mathrm{Re} \, \overline{\eta}(\omega) - \ci \, \mathrm{Im} \, \overline{\eta}(\omega). ~~~
\eea
It is straightforward to see that
\beaa \label{imager}
\mathrm{Im} \, \overline{\eta}(\omega) &= \frac{2}{\pi} \int_0^\infty d \omega' J(\omega') \int_0^\infty d t \sin \omega' t \sin \omega t ~~~
\\
& = \left\{ \begin{array}{lll} J(\omega) & \text{if} & \omega \geq 0 \\ - J(-\omega) & \text{if} & \omega < 0 \end{array} \right.,
\eeaa
and, through the Sokhotski–Plemelj theorem \cite{Bitsadze} (a.k.a. Kramers-Kronig relations),
\bea \label{chichi}
\chi(\omega) := \mathrm{Re} \, \overline{\eta}(\omega) = \frac{1}{\pi} \mathcal{P} \int_{-\infty}^\infty d \omega' \frac{J(\omega')}{\omega' - \omega},
\eea
where, and henceforth, it is understood that $J(\omega)$ is extended to negative arguments as an odd function, in accordance with Eq.~\eqref{imager}. The sign $\mathcal{P}$ signifies that one takes the Cauchy principal value of the integral, namely, $\mathcal{P} \int_{-\infty}^\infty := \lim\limits_{\epsilon \to 0} \Big[ \int_{- \infty}^{\omega - \epsilon} + \int_{\omega + \epsilon}^\infty \Big]$.

Note that, as long as $t_0 \to - \infty$, the only solution of the free equation \eqref{meq}, namely, one with $\xi(t)$ put to zero, is $q(t) \equiv 0$, which is a consequence of $\alpha(\omega) \neq 0$ for any value of $\omega$, provided that $J(\omega) = 0$ only for $\omega = 0$ and $\omega_0 >0$ (which is what we assume throughout this paper). Therefore, the unique solution of Eq.~\eqref{meq} for any initial conditions is
\beaa \label{qtpt}
q(t) = & \, \int_{-\infty}^\infty \frac{d \omega}{2 \pi} e^{- \ci \omega t} \frac{\overline{\xi}(\omega)}{\alpha(\omega)},
\\
p(t) = & \, - \ci \int_{-\infty}^\infty \frac{d \omega}{2 \pi} e^{- \ci \omega t} \frac{\omega \overline{\xi}(\omega)}{\alpha(\omega)}.
\eeaa
Note that, since $t_0 \to -\infty$, these solutions are the steady-state solutions (the time-dependence is simply because $[H_s^\CL, q] \neq 0$). Now, since the initial state of the bath is a Gibbs state, and therefore $\tr[p_k \tau_B] = \tr[q_k \tau_B] = 0$, we have
\bea
\av{q}_\infty \! := \tr \! \big[q \Omega_\infty^\CL \big] \! = \tr \! \big[q(t) \rho_0^\CL \! \otimes \tau_B^\CL \big] \! = 0 ~~~~~~
\eea
and, analogously,
\bea
\av{p}_\infty = 0.
\eea

Furthermore, taking into account that the initial state of the bath is a thermal state, and hence,
\bea \label{cnoise}
\langle \overline{\xi}(\omega) \overline{\xi}(\omega') \rangle = 2 \pi \delta(\omega + \omega') J(\omega) \coth \frac{\omega}{2 T},
\eea
it is easy to show that the steady-state covariance matrix $\sigma^\infty$ of the oscillator (see Eq.~\eqref{covmatdef}) is:
\bea \label{sgm11app}
\sigma_{11}^\infty := \langle q(t)^2 \rangle_{t \to \infty} &=& \frac{1}{\pi} \int_0^\infty d\omega \frac{J(\omega)}{|\alpha(\omega)|^2} \coth\frac{\omega}{2T},
\\ \label{sgm22app}
\sigma_{22}^\infty := \langle p(t)^2 \rangle_{t \to \infty} &=& \frac{1}{\pi} \int_0^\infty d\omega \frac{J(\omega) \omega^2}{|\alpha(\omega)|^2} \coth\frac{\omega}{2T}, ~~~~~
\eea
and $\sigma_{12}^\infty = \sigma_{21}^\infty := \tfrac{1}{2} \langle \{ q, p \} \rangle_{t \to \infty} = 0$.

Lastly, let us find $\av{\ddot{q} q}_\infty$ since we need it in Sec.~\ref{sec:CLenergetics}. Calculating $\ddot{q}$ using Eq.~\eqref{qtpt}, multiplying it by $q(t)$ from the same equation, and then using Eq.~\eqref{cnoise}, we find that
\bea
\av{\ddot{q} q}_\infty = - \frac{1}{\pi} \int_0^\infty \! d\omega \frac{J(\omega) \omega^2}{|\alpha(\omega)|^2} \coth\frac{\omega}{2T} = - \sigma_{22}^\infty. ~~~
\eea

\section{{GAUSSIAN ERGOTROPY}}
\label{app:GErgotropy}

Here we show how to calculate the maximal work cyclic Gaussian Hamiltonian processes can extract from a Gaussian state. We give this maximal amount of work the natural name of Gaussian ergotropy as all the involved states, Hamiltonians, and (therefore) unitaries are Gaussian.

\subsection{Setting notation}

Suppose we are given a $d$-mode bosonic system described by $d$ pairs of coordinates and momenta $(q_i,p_i)$. All Hamiltonians are going to be quadratic, so, composing the column vector $x=(q_1, p_1, \cdots, q_d, p_d)^T$, we introduce the symmetric matrix $M$ via
\bea
H = \frac{1}{2} x^T M x.
\eea
Note that $M$ has to be positive-semidefinite because otherwise $H$ will have a spectrum unbounded from below, which is unphysical. 

Let us furthermore define the covariance matrix, $\sigma$, as
\bea \label{covmatdef}
\sigma_{ij} = \frac{1}{2} \tr\left( \rho \{ x_i, x_j \} \right),
\eea
where $\rho$ is the state of the system (cf. Eq.~\eqref{lobzik}). Note that $\sigma$ is a symmetric, positive-definite matrix. Thus, for the average energy, we obtain
\bea \label{eq:avene}
\langle H \rangle := \tr(\rho H) = \frac{1}{2} \tr(\sigma M).
\eea

In order to proceed, we need to keep in mind the following two facts:

(i) Any Gaussian unitary transformation of the state taking $\rho$ to $\rho' = U \rho U^\dagger$ amounts to the covariance matrix $\sigma$ evolving into $\sigma' = \Lambda_U \sigma \Lambda_U^T$:
\bea \label{eq:unisym}
\rho \to U \rho U^\dagger \quad \Longleftrightarrow \quad \sigma' \to \Lambda_U \sigma \Lambda_U^T,
\eea
where $\Lambda_U$ is a real symplectic matrix; symplectic is any matrix satisfying $\Lambda_U J \Lambda_U^T = J$, where
\bea
J = \bigoplus_{i=1}^d J_1,
\eea
with
\bea
J_1 = \left(\begin{array}{cc}
 0 & 1 \\
-1 & 0
\end{array}\right),
\eea
is the symplectic identity; the real symplectic matrices constitute a group usually denoted by $\mathrm{Sp}(2 d, \mathbb{R})$. The statement in Eq.~\eqref{eq:unisym} works in both directions: any symplectic transformation of $\sigma$ can be generated by a Gaussian unitary evolution of $\rho$.

(ii) The Williamson's theorem \cite{Williamson_1936} holds. Namely, any $2d \times 2d$ symmetric matrix $\sigma \geq 0$ has $d$ nonnegative symplectic eigenvalues $s_1, \cdots, s_d$ and there exits a symplectic matrix $\Lambda_\sigma$ such that
\bea \label{eq:willy}
\sigma = \Lambda_\sigma s \Lambda_\sigma^T,
\eea
where
\bea \label{eq:nilly}
s = \bigoplus_{i=1}^d s_i \id_2,
\eea
with $\mathbb{I}_2$ denoting the $2 \times 2$ identity matrix. If we want $s_1 \leq \cdots \leq s_d$, then we add a $\uparrow$ superscript to $\Lambda_\sigma$ and $s$: $\sigma = \Lambda_\sigma^{\uparrow} s^\uparrow (\Lambda_\sigma^{\uparrow})^T$. If we want $s_1 \geq \cdots \geq s_d$, then $\sigma = \Lambda_\sigma^{\downarrow} s^\downarrow (\Lambda_\sigma^{\downarrow})^T$.

In Sec.~\ref{app:Willy} below we show how to obtain the Williamson decomposition from canonical Schur decomposition. The latter is built into most major packages (such as Mathematica, SciPy, ALGLIB, etc), thus the method provides a ready protocol for numerically computing the Williamson decomposition. Note that the presented method is a widely used one and is by no means an invention of ours.

\subsection{The ergotropy}

Gaussian ergotropy is the maximal amount of work cyclic Gaussian Hamiltonian processes can extract from a Gaussian state. Keeping in mind that any Gaussian Hamiltonian process generates a Gaussian unitary evolution operator and, vice versa, any Gaussian unitary evolution operator can be generated by a Gaussian Hamiltonian process, we can write the Gaussian ergotropy, $\mathcal{G}$, as
\bea \label{eq:Gerg1}
\mathcal{G} = \tr[\rho H] - \min_{U} \tr[U \rho U^\dagger H],
\eea
where the minimization is carried over the set of all Gaussian unitary operators. The state $U \rho U^\dagger$ delivering the minimum is called Gaussian-passive and such states were fully characterized in Ref.~\cite{Brown_2016}.

Now, recalling Eqs.~\eqref{eq:avene}, \eqref{eq:unisym}, and \eqref{eq:willy}, we can write
\bea \nonumber
\tr[U \rho U^\dagger H] &=& \frac{1}{2} \tr \left[\Lambda_U \Lambda_\sigma^\uparrow s^\uparrow (\Lambda_\sigma^\uparrow)^T \Lambda_U^T \Lambda_M^\downarrow m^\downarrow (\Lambda_M^\downarrow)^T \right]
\\ \nonumber
&=& \frac{1}{2} \tr \left[\Lambda s^\uparrow \Lambda^T m^\downarrow \right],
\eea
where $m^\downarrow$ is the symplectically diagonalized $M$ (with correspondingly ordered elements) (namely, denoting the symplectic eigenvalues of $M$ by $m_i$, $m^\downarrow = \bigoplus_{i = 1}^d m_i^\downarrow \id_2$) and $\Lambda = (\Lambda_M^\downarrow)^T \Lambda_U \Lambda_\sigma^\uparrow$ is a real symplectic matrix. 

Thus, we can rewrite Eq.~\eqref{eq:Gerg1} as
\bea \label{eq:Gerg2}
\mathcal{G} = \frac{1}{2} \tr[\sigma M] - \frac{1}{2} \min_{\Lambda J \Lambda^T = J} \tr \left[ \Lambda s^\uparrow \Lambda^T m^\downarrow \right].
\eea
Introducing $\{ S_n(\Lambda) \}_{n=0}^d$ such that
\bea \nonumber
S_0(\Lambda) = 0, \;\, S_k (\Lambda) = \sum_{i=1}^{2k} \left( \Lambda s^\uparrow \Lambda^T \right)_{ii}, \;\, \mathrm{for} \;\, k = \overline{1,d}, ~
\eea
we have
\beaa \label{eq:Abel}
\tr \! \left[ \Lambda s^\uparrow \Lambda^T m^\downarrow \right] \! =& \sum_{k=1}^d \! m_{k}^\downarrow \left[ S_{k}(\Lambda)-S_{k-1}(\Lambda) \right]
\\
=& \sum_{k=1}^{d-1} \! \big[ m_{k}^\downarrow - m_{k+1}^\downarrow \big] S_k(\Lambda) + m_{d}^\downarrow S_d(\Lambda), ~~~~
\eeaa
whence it immediately follows that
\beaa \nonumber
\min_{\Lambda} \tr \left[ \Lambda s^\uparrow \Lambda^T m^\downarrow \right] \geq \sum_{k=1}^{d - 1} \big[ m_{k}^\downarrow - m_{k + 1}^\downarrow \big] & \min_{\Lambda} S_k(\Lambda)
\\
+& \; m_{d}^\downarrow \min_{\Lambda} S_d(\Lambda), ~~
\eeaa
where the minimization is carried over the whole group of real symplectic matrices, i.e., $\Lambda \in \mathrm{Sp}(2 d, \mathbb{R})$. Note that there is no a priori guarantee that a single $\Lambda$ will minimize all $S_k(\Lambda)$'s; however, invoking Lemma 1 of Ref.~\cite{Hiroshima_2006}, stating that
\bea
\min_{\Lambda} S_k(\Lambda) = 2 \sum_{i = 1}^k s_i^\uparrow,
\eea
we see that $\Lambda = \id_{2d}$ simultaneously minimizes all $S_k(\Lambda)$'s, thereby showing that
\bea
\min_{\Lambda} \tr \left[ \Lambda s^\uparrow \Lambda^T m^\downarrow \right] = 2 \sum_{k=1}^d s_k^\uparrow m_k^\downarrow.
\eea
We have thus just proven the main result of this section, namely, that
\bea \nonumber
\mathcal{G} = \frac{1}{2} \tr[\sigma M] - \frac{1}{2} \tr[s^\uparrow m^\downarrow] = \frac{1}{2} \tr[\sigma M] - \sum_{k=1}^d s_k^\uparrow m_k^\downarrow,
\\ \label{eq:Gerg3}
\eea
and the maximum is delivered by the unitary $U$ that generates the real symplectic transformation
\bea \label{eq:GergUni}
\Lambda_U = - J \Lambda_M^\downarrow (\Lambda_\sigma^\uparrow)^T J.
\eea
To the best of our knowledge, Eq.~\eqref{eq:Gerg3} in this general form has never been reported in the literature. An analogue of Eq.~\eqref{eq:Gerg3} for free-fermionic systems is reported in Ref.~\cite{Perarnau-Llobet_2016}.

It is a trivial consequence of Eq.~\eqref{eq:Gerg3} that, when the system consists of $d$ noninteracting modes, the Gaussian operation extracting maximal work is the one that evolves the initial state into $s^\uparrow$, if $M = m^\downarrow$, or $s^\downarrow$, if $M = m^\uparrow$. Note that, although the maximal work is always delivered by the symplectic transformation in Eq.~\eqref{eq:GergUni}, in some cases, it will not be unique. Indeed, as can be observed by inspecting Eq.~\eqref{eq:Abel}, when some of the normal frequencies coincide, the symplectic transformation of the state delivering the minimal final energy might not be unique. In fact, for $d=2$ and $m_1 = m_2$ (i.e., $m^\downarrow = m^\uparrow = m$), it was shown in Ref.~\cite{Brown_2016} that the minimum of $\tr[\Lambda \sigma \Lambda^T m]$ is delivered by both $\Lambda \sigma \Lambda^T = \sigma^\uparrow$ and $\Lambda \sigma \Lambda^T = s_\mathrm{canon}$, where $s_{\mathrm{canon}}$ is in the ``canonical'' form \cite{Simon_2000, Duan_2000}:
\bea
s_\mathrm{canon} = \left( \begin{array}{cc} a \id_2 & c \id_2 \\ c \id_2 & b \id_2 \end{array} \right).
\eea
Importantly, in the canonical form, the nondiagonal block matrices are generally of the form $\mathrm{diag}(c_1, c_2)$, whereas for $s_\mathrm{canon}$ above to deliver the minimum of energy it is necessary that $c_1 = c_2$ \cite{Brown_2016}. We elaborate on this in Sec.~\ref{app:canon} below.

\subsection{Ergotropy of a single oscillator}
\label{app:Gerg1osc}

Let us illustrate the above theory on the simple example of a single oscillator with Hamiltonian
\bea
H = \frac{p^2}{2} + \frac{\omega_0^2 q^2}{2} = \frac{1}{2} x^{\mathrm{T}} M_1 x,
\eea
where $x = \Big( \! \begin{array}{c} q \\ p \end{array} \! \Big)$ and
\bea
M_1 = \left(\begin{array}{cc} \omega_0^2 & 0 \\ 0 & 1 \end{array} \right),
\eea
that starts in a Gaussian state $\rho$ described by some $\sigma = \left(\begin{array}{cc} \sigma_{11} & \sigma_{12} \\ \sigma_{12} & \sigma_{22} \end{array} \right)$ (note that here we explicitly take into account that $\sigma$ is a symmetric matrix). 

Now, in order to use Eq.~\eqref{eq:Gerg3}, let us find the symplectic eigenvalues of $\sigma$ and $M_1$. Using the observation below Eq.~\eqref{dages}, we immediately find the (only one) symplectic eigenvalue of $\sigma$ to be $s_1 = \sqrt{\det \sigma}$ and that of $M_1$: $m_1 = \omega_0$. Eq.~\eqref{eq:Gerg3} then implies
\bea \label{1oscerg}
\mathcal{G}_1 = \frac{1}{2}(\sigma_{11} \omega_0^2 + \sigma_{22}) - \omega_0 \sqrt{\sigma_{11}\sigma_{22} - \sigma_{12}^2}.
\eea
Note that the passive state in which the system ends up, as a result of the ergotropy extraction, is the state which, when $H = \frac{\omega_0 P^2}{2}+ \frac{\omega_0 Q^2}{2}$, has a covariance matrix $s^\uparrow = \Big( \! \begin{array}{cc} s_1 & 0 \\ 0 & s_1 \end{array} \! \Big)$, which means that the final passive state, $\rho_p$, is a Gibbs state at some temperature: $\rho_p \propto e^{- \beta_p H}$, where the temperature can be determined from $S(\rho_p) = S(\rho)$. It obviously follows from this that the Gaussian ergotropy coincides with the full ergotropy (i.e., one found by optimizing over all---not only Gaussian---unitary operations), as the Gibbs state has the lowest energy for a given value of entropy. We note that the formula \eqref{1oscerg} can be deduced from the analysis in Ref.~\cite{Brown_2016}; see also Ref.~\cite{Farina_2019}.

Note also that, upon bringing $H$ back to the $\frac{p^2}{2} + \frac{\omega_0^2 q^2}{2}$ form, i.e., switching to
\bea
\Big( \begin{array}{c} q \\ p \end{array} \Big) = \left( \begin{array}{cc} \frac{1}{\sqrt{\omega_0}} & 0 \\ 0 & \sqrt{\omega_0} \end{array} \right) \Big( \begin{array}{c} Q \\ P \end{array} \Big),
\eea
the covariance matrix of the final (Gaussian-)passive state will take the form
\bea \label{borjomi}
\sigma_p = \sqrt{\sigma_{11} \sigma_{22} - \sigma_{12}^2} \left( \begin{array}{cc} \frac{1}{\omega_0} & 0 \\ 0 & \omega_0 \end{array} \right).
\eea

The temperature $T_p = \beta_p^{-1}$ can be determined by noting that the covariance matrix elements of a free harmonic oscillator in a Gibbs state are given by Eq.~\eqref{freeosc}, therefore, comparing with Eq.~\eqref{borjomi}, we find
\bea \nonumber
\frac{1}{\omega_0}\sqrt{\sigma_{11} \sigma_{22} - \sigma_{12}^2} &=& \frac{1}{2 \omega_0} \coth\frac{\omega_0}{2 T_p},
\\ \nonumber
&\Downarrow&
\\
\beta_p = \frac{2}{\omega_0} \mathrm{arccoth} \!\!\!\! && \!\!\!\! \p{2 \sqrt{\sigma_{11} \sigma_{22} - \sigma_{12}^2}}.
\eea

\subsection{Technical nuances}

\subsubsection{Williamson decomposition from Schur decomposition}
\label{app:Willy}

Let us describe how, given a $\sigma$, to obtain the $\Lambda$ and the $s$ in Eq.~\eqref{eq:willy} numerically. To that end, we are going to make use of the standard Schur decomposition for skew-symmetric matrices: Say, $A$ is a $2d \times 2d$ real skew-symmetric matrix (i.e., $A^T=-A$), then there exists a real orthogonal matrix $O$ such that
\bea \label{eq:Schur1}
A = O \alpha O^T,
\eea
with
\bea \label{eq:Schur2}
\alpha = \bigoplus_{i=1}^d a_i J_1 = a^{1/2} J a^{1/2},
\eea
where $a_i$ are real, nonnegative numbers and
\bea
a = \bigoplus_{i=1}^d a_i \mathbb{I}_2.
\eea
This decomposition is a built-in function in most numerical software packages. Notice that, as Eqs.~\eqref{eq:Schur1} and \eqref{eq:Schur2} readily suggest, $\pm \mathrm{i} a_i$ are the eigenvalues of $A$.

Now, noticing that the matrix $\sigma^{1/2} J \sigma^{1/2}$ is skew-symmetric, let us write its Schur decomposition as
\bea
\sigma^{1/2} J \sigma^{1/2} = O s^{1/2} J s^{1/2} O^T.
\eea
From there, we immediately see that $\bar{\Lambda} = \sigma^{-1/2} O s^{1/2}$ is a symplectic matrix and that $\bar{\Lambda}^T \sigma \bar{\Lambda} = s$. Introducing the symplectic matrix
\bea
\Lambda_\sigma = \left(\bar{\Lambda}^T \right)^{-1} = -J \bar{\Lambda} J = -J \sigma^{-1/2} O s^{1/2} J, ~~
\eea
we observe that
\bea \label{dages}
\sigma = \Lambda_\sigma s \Lambda_\sigma^T.
\eea
Moreover, $\pm \mathrm{i} s_i$ are the eigenvalues of $\sigma^{1/2} J \sigma^{1/2}$, or, equivalently, of $\sigma J$.

\subsubsection{Canonical form vs Williamson form} \label{app:canon}

Say, we have two noninteracting modes with both frequencies equal to some $\omega_0$ (i.e., $m_1 = m_2 = \omega_0$). Let us now see how
\bea \label{eq:B1}
E_\mathrm{canon} = \frac{1}{2} \tr(s_\mathrm{canon} m) = \omega_0 (a + b),
\eea
where we keep $c_1$ and $c_2$ general, compares to
\bea \label{eq:B2}
E_\mathrm{min} = \frac{1}{2} \tr(s^\uparrow m) = \omega_0 (s_1 + s_2).
\eea
First of all, we note that the covariance matrix is a positive-semidefinite matrix, which is equivalent to
\bea
c_1^2 \leq ab \quad \mathrm{and} \quad c_2^2 \leq ab.
\eea
Next, it is easy to calculate the symplectic eigenvalues of $s_\mathrm{canon}$, thereby relating $s_1$ and $s_2$ with $a$, $b$, $c_1$, and $c_2$:
\bea
s_1 &=& \sqrt{\frac{a^2 + b^2 + 2 c_1 c_2 - \kappa}{2}},
\\
s_2 &=& \sqrt{\frac{a^2 + b^2 + 2 c_1 c_2 + \kappa}{2}},
\eea
where
\bea \nonumber
\kappa = \sqrt{(a^2 - b^2)^2 + 4 (a^2 + b^2) c_1 c_2 + 4 a b (c_1^2 + c_2^2)}. ~~
\eea
Now, it is straightforward to see that
\bea \nonumber
(s_1 + s_2)^2 &=& a^2 + b^2 + 2 c_1 c_2 + 2 \sqrt{(a b - c_1^2) (a b - c_2^2)}, ~
\eea
where, using the inequality $2\sqrt{xy}\leq x +y $ on the square-root term, we obtain
\bea
s_1 + s_2 \leq \sqrt{(a+b)^2 - (c_1 - c_2)^2},
\eea
with an equality if and only if $c_1 = c_2$, in which case one obtains $s_1 + s_2 = a + b$, which, in turn (cf. Eqs.~\eqref{eq:B1} and \eqref{eq:B2}) implies $E_\mathrm{canon} = E_\mathrm{min}$, meaning that the canonical form with (and only with) $c_1 = c_2$ does indeed have minimal energy, corroborating Theorem 1 of Ref.~\cite{Brown_2016}.

\section{{HIGH-TEMPERATURE LIMIT IN THE CALDEIRA-LEGGETT MODEL: ASYMPTOTIC EQUIPARTITION}}
\label{app:sigmahighT}

Here we analyze the behavior of $\sigma^\infty$ when $T \gg \omega_0$. Inspecting Eq.~\eqref{sgmii}, we immediately notice that, to the integrals, only those $\omega$'s contribute for which $\omega/T < 1$, as for $\omega > T$ the integrand is suppressed by strongly decaying $f(\omega/\omega_c)$ and large $\alpha(\omega)$. With this and the expansion $\coth x = \frac{1}{x} + \frac{x}{3} + O (x^3)$, $x \gg 1$, in mind, in the $T \gg \omega_0$ limit, we can write
\bea
\sigma^\infty_{11} &=& T \iota_1 + O (1 / T),
\\
\sigma^\infty_{22} &=& T \iota_2 + O (\omega^2_0 / T),
\eea
where
\bea \label{iota1}
\iota_1 &=& \frac{2}{\pi} \int_0^\infty d\omega \frac{J(\omega)}{\omega |\alpha(\omega)|^2},
\\ \label{iota2}
\iota_2 &=& \frac{2}{\pi} \int_0^\infty d\omega \frac{\omega J(\omega)}{|\alpha(\omega)|^2}.
\eea

In order to calculate $\iota_1$, keeping in mind Eqs.~\eqref{alfombra} and \eqref{imager}, we observe that
\bea \label{alfalfa}
\frac{J(\omega)}{|\alpha(\omega)|^2} = \mathrm{Im} \frac{1}{\alpha(\omega)}
\eea
and note that, as long as $J(\omega)$ is analytic in the closed upper half plane, so is $\frac{1}{\alpha(\omega)}$. Moreover, in the upper half plane, $\alpha(\omega)$ decays faster than $1/|\omega|$ (it is $\propto |\omega|^{-2}$) for $|\omega| \gg \omega_0$. Therefore, the Kramers-Kronig formulas hold for $1/\alpha(\omega)$:
\bea
\mathrm{Re} \frac{1}{\alpha(\omega)} = \frac{1}{\pi} \mathcal{P} \int_{-\infty}^\infty d\omega' \frac{1}{\omega' - \omega} \mathrm{Im} \frac{1}{\alpha(\omega')}.
\eea
Here, choosing $\omega = 0$, noting that $\alpha(-\omega) = \alpha(\omega)$, using the identity \eqref{alfalfa}, and further noting that the antiderivative of $J(\omega)/\omega$ is regular as $\omega \to 0$ (which is the case as long as, for $\omega \to 0$, $J(\omega) \propto \omega^\varsigma$, with some $\varsigma > 0$, which is the case in this paper), we find
\bea
\mathrm{Re} \frac{1}{\alpha(0)} = \frac{2}{\pi} \int_0^\infty d\omega \frac{J(\omega)}{\omega |\alpha(\omega)|^2}.
\eea
Lastly, noting that $\alpha(0) = \omega_0^2 + \omega_R^2 - \chi(0)$, and, remembering Eqs.~\eqref{chichi} and \eqref{omerta}, which tell us that $\chi(0) = \omega^2_R$, we find that $\alpha(0) = \omega_0^2$. In other words:
\bea \label{beauty1}
\frac{2}{\pi} \int_0^\infty d\omega \frac{J(\omega)}{\omega |\alpha(\omega)|^2} = \frac{1}{\omega_0^2}.
\eea

With the other integral, $\iota_2$, the same argument using the Kramers-Kronig relations will not work. Indeed, the integrand in Eq.~\eqref{iota2} is $\omega/\alpha(\omega)$, which, albeit analytic in the closed upper half plane, is $\propto 1/|\omega|$ for $\omega \gg \omega_0$, where, for the Kramers-Kronig relations to work, it should have decayed strictly faster than $1/|\omega|$ \cite{Bitsadze}. Instead, by an explicit calculation, we will show that $\iota_2$ is equal to $\frac{1}{\ci} [q(t), p(t)] = 1$. (That $[q(t), p(t)] = \ci$, for any $t$, is a trivial consequence of the fact that the global system-bath evolution, generated by $H_\gT^\CL$, is unitary.) Indeed, using Eq.~\eqref{qtpt}, it is a simple exercise to arrive to
\bea \nonumber
[q(t), p(t)] = \frac{1}{4 \pi^2} \int_{-\infty}^\infty d\omega_1 d\omega_2 (-\ci \omega_2) \frac{[\overline{\xi}(\omega_1), \overline{\xi}(\omega_2)]}{\alpha(\omega_1) \alpha(\omega_2)}.
\\ \label{kyazh}
\eea
On the other hand, using Eq.~\eqref{zhesht}, it is easy to show that
\bea
[\xi(t_2), \xi(t_1)] = \frac{2 \ci}{\pi} \int_0^\infty d\omega J(\omega) \sin \omega(t_1 - t_2), ~~
\eea
therefore,
\beaa \nonumber
[\overline{\xi} & (\omega_2), \overline{\xi}(\omega_1)] = \int_{-\infty}^\infty dt_1 dt_2 \, e^{\ci \omega_2 t_2 + \ci \omega_1 t_1} [\xi(t_2), \xi(t_1)] 
\\
&= 4 \pi \delta(\omega_1 + \omega_2) \int_0^\infty d\omega J(\omega) [\delta(\omega - \omega_2) - \delta(\omega - \omega_1)], ~
\eeaa
which, substituted into Eq.~\eqref{kyazh}, gives
\bea
[q(t), p(t)] = \frac{2 \ci}{\pi} \int_0^\infty d\omega \frac{\omega J(\omega)}{|\alpha(\omega)|^2}.
\eea
Taking $[q(t), p(t)] = \ci$ into account, we thus find that
\bea \label{beauty2}
\frac{2}{\pi} \int_0^\infty d\omega \frac{\omega J(\omega)}{|\alpha(\omega)|^2} = 1.
\eea

The obtained formulas can be cast in the form
\bea \label{equipart:q}
\av{\frac{\omega_0^2 q^2}{2}}_{t \to \infty} &=& \frac{T}{2} + O \bigg( \frac{\omega^2_0}{T} \bigg),
\\ \label{equipart:p}
\av{\frac{p^2}{2}}_{t \to \infty} &=& \frac{T}{2} + O \bigg( \frac{\omega^2_0}{T} \bigg),
\eea
which clearly shows that, in the high-temperature limit, the energy per each canonical variable is $\approx T/2$. This situation is met in classical statistical mechanics, where it is referred to as equipartition theorem \cite{ll5old}. Obviously, Eqs.~\eqref{equipart:q} and \eqref{equipart:p} also hold for a free oscillator (i.e., one that is not coupled to an environment, but $\omega_0 >0$).

See Ref.~\cite{Bialas_2018} for an alternative proof of Eqs.~\eqref{beauty1} and \eqref{beauty2}. See also Ref.~\cite{Philbin_2016} where a similar result, albeit in a slightly different setting, was obtained.

\section{{ASYMPTOTIC EXPANSION OF THE COVARIANCE MATRIX WITH RESPECT TO THE COUPLING CONSTANT}}
\label{app:weaksigma}

In the subsequent subsections, we will derive the weak coupling expansion of the covariance matrix of an oscillator coupled to a Caldeira-Leggett bath and use that expansion to calculate the figures of merit of the device. We will also explore the low-temperature limit.

In the weak coupling limit, one expects the steady state of the system to be close to the thermal equilibrium state ($\tau_s^\CL$), characterized by the covariance matrix comprised by $\sigma^{\mathrm{(free)}}_{11}$ and $\sigma^{\mathrm{(free)}}_{22}$ in Eq.~\eqref{freeosc}. Below, we will see that this is indeed the case and, moreover, will find the $O (\gamma)$ correction to $\sigma^{\mathrm{(free)}}_{11}$ and $\sigma^{\mathrm{(free)}}_{22}$ for $\gamma \ll 1$.

\subsection{Asymptotic expansion of \texorpdfstring{$\sigma_{11}$}{sigma11} with respect to \texorpdfstring{$\gamma \ll 1$}{gamma<<1}}
\label{app:sigma1}

Performing the following change of variable in the integral in Eq.~\eqref{sgmii} for $\sigma_{11}^\infty$:
\bea
x = 1 - \left(\frac{\omega}{\omega_0}\right)^2, \qquad x \in (-\infty, 1],
\eea
invoking Eq.~\eqref{J:def}, and noticing that $\omega_R^2 - \chi(\omega)$ is linearly proportional to $\gamma$, and therefore
\bea \label{ggg}
g(x) = \frac{\omega_R^2 - \chi(\omega_0 \sqrt{1 - x})}{\gamma \omega_0^2}
\eea
is independent on $\gamma$, we find
\bea \nonumber
\sigma_{11} = \frac{1}{2 \omega_0 \pi \gamma} \int_{-\infty}^1 dx \frac{F(x) \coth \left( \frac{\omega_0}{2 T} \sqrt{1 - x} \right)}{\big[ \frac{x}{\gamma} + g(x) \big]^2 + (1 - x) F(x)^2}, ~~~
\\ \label{sigma1}
\eea
where, for convenience, we have introduced
\bea \label{FK0}
F(x) = f \left( \omega_0 \omega_c^{-1}\sqrt{1 - x} \right).
\eea
In Eq.~\eqref{sigma1}, the dependence of $\sigma_{11}$ on $\gamma$ is localized in one place. Inspecting the part of the integrand without the $\coth$, we immediately recognize a similarity to the
\bea \label{DDF}
\lim\limits_{\gamma \to 0} \frac{1}{\pi} \frac{\gamma}{z^2 + \gamma^2} = \delta(z)
\eea
representation of Dirac's delta function, and thus expect the leading term in Eq.~\eqref{sigma1} to be $ \propto \frac{1}{2 \omega_0} \coth \frac{\omega_0}{2 T}$, which is what one should indeed obtain (cf. Eq.~\eqref{freeosc}). However, we are going to need the higher-order terms in the expansion of $\sigma_{11}$ around $\gamma = 0$, and in order to make further progress, we note that the behavior of the integrand in Eq.~\eqref{sigma1} behaves differently when $x \propto \gamma$ (it is finite) and when $x \propto 1$ (it tends to zero with $\gamma \to 0$). To isolate different behaviors, we divide the integration region as
\bea \label{divide}
\int_{-\infty}^1 = \underbrace{\int_{-\infty}^{-\gamma^{1/3}}}_{I_1} + \underbrace{\int_{-\gamma^{1/3}}^{\gamma^{1/3}}}_{I_2} + \underbrace{\int_{\gamma^{1/3}}^1}_{I_3}.
\eea

Let us first deal with $I_1$. Introducing, for convenience,
\bea \label{FK1}
K(x) = F(x) \coth \Big(\frac{\omega_0}{2 T} \sqrt{1 - x} \Big),
\eea
and further denoting
\bea \label{FK2}
\widetilde{F}(x) = F(-x) \quad \text{and} \quad \widetilde{K}(x) = K(-x),
\eea
we obtain
\bea \label{I1}
I_1 = \int_{\gamma^{1/3}}^\infty dx \frac{\widetilde{K}(x)}{\big[ \frac{x}{\gamma} - g(-x) \big]^2 + (1 + x) \widetilde{F}(x)^2}.
\eea
Keeping in mind that $\widetilde{F}(x)$ and $\widetilde{K}(x)$ are quickly-decaying functions for $x > (\omega_c/\omega_0)^2$, we do not concern ourselves with the large-$x$ behavior of $g(x)$ (which, for some generic choices of $f$, is a low-degree polynomial of $x$) and consider it small (``finite'') as compared to $x/\gamma$ in the asymptotic limit of $\gamma \to 0$. Expanding the integrand in Eq.~\eqref{I1} around $\gamma = 0$, we find
\bea \label{razite}
I_1 = \gamma^2 \int_{\gamma^{1/3}}^\infty dx \frac{\widetilde{K}(x)}{x^2} + 2\gamma^3 \int_{\gamma^{1/3}}^\infty dx \frac{\widetilde{K}(x) g(-x)}{x^3} ~~~~~~~
\eea
plus higher-order terms. Noting that, for $k > 1$,
\bea
\int_{\gamma^{1/3}} dx \frac{\text{regular function}}{x^k} \propto \gamma^{-\frac{k-1}{3}},
\eea
we see that the first term in Eq.~\eqref{razite} scales as $\gamma^{5/3}$ and the second term scales as $\gamma^{7/3}$, meaning that, to $\sigma_{11}$, these contribute as $\gamma^{2/3}$ and $\gamma^{4/3}$, and, since we are interested in the next to the leading order term in the expansion of $\sigma_{11}$ with respect to $\gamma$, we will discard the second term and focus on the first one:
\bea \nonumber
I_1 = \gamma^2 \int_{\gamma^{1/3}}^A dx \frac{\widetilde{K}(x)}{x^2} + \gamma^2 \int_{A}^\infty dx \frac{\widetilde{K}(x)}{x^2} + O \big( \gamma^{7/3} \big).
\eea
Now, since $\widetilde{K}(x)$ is a regular, analytic function, we can write
\bea \nonumber
\frac{\widetilde{K}(x)}{x^2} = \frac{\widetilde{K}(0)}{x^2} + \frac{\widetilde{K}'(0)}{x} + \sum_{k=2}^\infty \frac{\widetilde{K}^{(k)}(0)}{k!} x^{k-2}.
\eea
Hence,
\beaa \nonumber
I_1 =& \, \gamma^{5/3} \widetilde{K}(0) + \gamma^2 \ln \! \bigg( \frac{A}{\gamma^{1/3}} \bigg) \widetilde{K}'(0) + \gamma^2 \! \int_{A}^\infty \! dx \frac{\widetilde{K}(x)}{x^2}
\\
&- \gamma^2 \frac{\widetilde{K}(0)}{A} + \gamma^2 \sum_{k=2}^\infty \frac{\widetilde{K}^{(k)}(0)}{k!}\frac{A^{k-1}}{k-1} + O \big( \gamma^{7/3} \big).
\eeaa
We further notice that
\beaa \nonumber
\sum_{k=2}^\infty \frac{\widetilde{K}^{(k)}(0)}{k!}\frac{A^{k - 1}}{k - 1} &= \frac{1}{A} \sum_{k = 2}^\infty \frac{k \widetilde{K}^{(k)}(0)}{(k - 1) k!}\frac{A^k}{k}
\\
=& \, \frac{1}{A} \sum_{k = 2}^\infty \frac{\widetilde{K}^{(k)}(0)}{(k - 1)(k - 1)!} \int_0^A da a^{k - 1}
\\
=& \, \frac{1}{A} \sum_{k = 1}^\infty \frac{\widetilde{K}^{(k + 1)}(0)}{k!} \int_0^A da \int_0^a db b^{k - 1} ~
\\
=& \, \frac{1}{A} \int_0^A da \int_0^a \frac{db}{b} \sum_{k = 1}^\infty \frac{\widetilde{K}^{(k + 1)}(0)}{k!} b^k
\\
=& \, \frac{1}{A} \int_0^A da \int_0^a db \frac{\widetilde{K}'(b) - \widetilde{K}'(0)}{b}.
\eeaa
So, setting $A = 1$, and noticing that $\widetilde{K}(0) = K(0)$ and $\widetilde{K}'(x) = - K'(-x)$, we obtain
\beaa \nonumber
I_1 =& \, \gamma^{5/3} K(0) - \gamma^2 \ln \bigg( \! \frac{1}{\gamma} \! \bigg) \frac{K'(0)}{3}
\\
&+ \gamma^2 \int_{-\infty}^{-1} dx \frac{K(x)}{x^2} - \gamma^2 K(0)
\\
&+ \gamma^2 \int_0^1 da \int_0^a db \frac{K'(0) - K'(-b)}{b} + O \big( \gamma^{7/3}\big). ~
\eeaa
By taking identical steps for $I_3$, we can immediately write
\beaa \nonumber
I_3 =& \, \gamma^{5/3} K(0) + \gamma^2 \ln \bigg( \! \frac{1}{\gamma} \! \bigg) \frac{K'(0)}{3} - \gamma^2 K(0)
\\
&+ \gamma^2 \int_0^1 da \int_0^a db \frac{K'(b) - K'(0)}{b} + O \big( \gamma^{7/3} \big),
\eeaa
to finally arrive at
\beaa \nonumber
I_1 + I_3 =& \, 2\gamma^{5/3} K(0) - 2 \gamma^2 K(0) + \gamma^2 \int_{-\infty}^{-1} dx \frac{K(x)}{x^2}
\\
&+ \gamma^2 \int_0^1 da \int_0^a db \frac{K'(b) - K'(-b)}{b} + O \big( \gamma^{7/3} \big).
\eeaa
Turning to $I_2$,
\bea
I_2 = \int_{-\gamma^{1/3}}^{\gamma^{1/3}} dx \frac{K(x)}{\left[ \frac{x}{\gamma} + g(x) \right]^2 + (1-x) F(x)^2}, ~~
\eea
let us switch the integration variable to $x/\gamma$, so that
\beaa \label{tutak}
I_2 &= \gamma \int_{-\gamma^{-2/3}}^{\gamma^{-2/3}} dx \frac{K(\gamma x)}{\left[ x + g(\gamma x) \right]^2 + (1 - \gamma x) F(\gamma x)^2} ~~~~~~~
\\
&:= \gamma \int_{-\gamma^{-2/3}}^{\gamma^{-2/3}} dx \nu_\gamma(x),
\eeaa
and notice that $\gamma x \ll 1$ in the whole integration interval, which allows us Taylor-expand the integrand around $\gamma x$ (while not touching $x$):
\beaa \nonumber
\nu_\gamma(x) =& \, \frac{K_0}{[x + g_0]^2 + F_0^2} + \bigg( \frac{x K'_0}{[x + g_0]^2 + F_0^2}
\\
&- \frac{x K_0 (2 x g'_0 + 2 F_0 F'_0 + 2 g_0 g'_0 - F_0^2)}{([x + g_0]^2 + F_0^2)^2} \bigg) \gamma + O \big( \gamma^2 \big),
\eeaa
where, for simplicity, we denoted
\beaa \nonumber
F_0 &:= F(0), \quad K_0 := K(0), \quad g_0 := g(0),
\\
F'_0 &:= F'(0), \quad K'_0 := K'(0), \quad g'_0 := g'(0).
\eeaa
Now, performing the integration in Eq.~\eqref{tutak} and denoting
\bea
L := \gamma^{-2/3},
\eea
and keeping in mind that $L \gg 1$, we obtain:
\beaa \nonumber
\frac{I_2}{\gamma} =& \, \frac{K_0}{F_0} \bigg[ \pi - \arctan \frac{2 L F_0}{L^2 - \lambda} \bigg] + \bigg( K'_0 \, \mathrm{arctanh} \frac{2 L g_0}{L^2 + \lambda}
\\
&+ \frac{L K_0}{F_0} \frac{2 F_0 g'_0 (L^2 + \lambda) - g_0 (F_0 - 2 F'_0) (L^2 - \lambda)}{L^4 + 2 L^2 (F_0^2 - g_0^2) + \lambda^2} ~~~
\\
&- \bigg[ \pi - \arctan \frac{2 L F_0}{L^2 - \lambda} \bigg] \! \cdot \! \bigg[ \frac{Y_0}{2} + Y'_0 \bigg] \bigg) \gamma + O (\gamma^2),
\eeaa
where we have introduced
\bea
\lambda := F_0^2 + g_0^2
\eea
and
\bea \nonumber
Y(x) = \frac{K(x) g(x)}{F(x)}, \quad \text{with} \quad Y_0 := Y(0), \;\, Y'_0 := Y'(0).
\\ \label{Yg}
\eea
Next, Taylor-expanding these expressions around $\frac{1}{L}=0$, we obtain
\beaa \label{I2mess4}
\frac{I_2}{\gamma} =& \pi \frac{K(0)}{F(0)} - 2 K(0) \gamma^{2/3} - \frac{\pi}{2} \left[ Y(0) + 2 Y'(0) \right] \gamma ~~~~~~
\\
&+ 4[K'(0) g(0) + K(0) g'(0)] \gamma^{5/3} + O (\gamma^2),
\eeaa
which means that
\beaa \nonumber
I_1 + I_2 + I_3 =& \, \gamma \pi \frac{K(0)}{F(0)} + \gamma^2 \! \int_0^1 \! da \! \int_0^a \! db \frac{K'(b) - K'(-b)}{b}
\\
&- 2 \gamma^2 K(0) - \gamma^2 \frac{\pi}{2} \left[ Y(0) + 2 Y'(0) \right]
\\
&+ \gamma^2 \int_{-\infty}^{-1} dx \frac{K(x)}{x^2} + O \big( \gamma^{7/3} \big),
\eeaa
and hence, keeping in mind Eqs.~\eqref{FK0}, \eqref{FK1}, and \eqref{FK2}, we find that 
\bea \label{sgm11asympt}
\sigma_{11} = \frac{1}{2 \omega_0} \coth\frac{\omega_0}{2 T} + \frac{\Phi_T}{2 \pi \omega_0}\gamma + \frac{1}{\omega_0} \circ (\gamma),
\eea
where
\beaa \label{Phi}
\Phi_T =& \int_{-\infty}^{-1} dx \frac{K(x)}{x^2} - 2 K(0) - \frac{\pi}{2} Y(0) - \pi Y'(0) ~~~~~~
\\
&+ \int_0^1 da \int_0^a db \frac{K'(b) - K'(-b)}{b}.
\eeaa
Recall that $F(x)$, $K(x)$, $Y(x)$, and $g(x)$ are defined in Eqs.~\eqref{FK0}, \eqref{FK1}, \eqref{Yg}, and \eqref{ggg}. From these equations, we find
\beaa \label{tashaxust}
K(0) &= f(\omega_0/\omega_c) \coth\frac{\omega_0}{2T}
\\
Y(0) &= \frac{\omega_R^2 - \chi (\omega_0)}{\gamma \omega_0^2} \coth \frac{\omega_0}{2 T}
\\
Y'(0) &= \frac{\chi'(\omega_0)}{2 \gamma \omega_0} \coth\frac{\omega_0}{2 T} + \frac{\omega_R^2 - \chi (\omega_0)}{4 \gamma \omega_0 T} \frac{1}{\sinh^2 \frac{\omega_0}{2 T}}. ~~~~
\eeaa

\subsection{Asymptotic expansion of \texorpdfstring{$\sigma_{22}$}{sigma2} with respect to \texorpdfstring{$\gamma \ll 1$}{gamma<<1}}
\label{app:sigma2}

Turning to $\sigma_{22}$ (see Eq.~\eqref{sgmii}), we immediately see that its analysis is identical to that for $\sigma_{11}$, only multiplied by $\omega_0^2$ and with $K(x)$ substituted by $\overline{K}(x) = (1 - x) K(x)$ (since $\omega^2 = \omega_0^2 (1 - x)$). Thus, from Eqs.~\eqref{sgm11asympt} and \eqref{Phi}, we read:
\bea \label{sgm22asympt}
\sigma_{22} = \frac{\omega_0}{2} \coth\frac{\omega_0}{2 T} + \frac{\omega_0 \Psi_T}{2 \pi}\gamma + \omega_0 \circ (\gamma),
\eea
where
\beaa \nonumber
\Psi_T =& \int_{-\infty}^{-1} dx \frac{\overline{K}(x)}{x^2} - 2 \overline{K}(0) - \frac{\pi}{2} \overline{Y}(0) - \pi \overline{Y}'(0)
\\
&+ \int_0^1 da \int_0^a db \frac{\overline{K}'(b) - \overline{K}'(-b)}{b}.
\eeaa
Substituting $\overline{K}(x) = (1 - x) K(x)$ here and in Eq.~\eqref{Yg}, we obtain 
\beaa \label{Psi}
\Psi_T =& \int_{-\infty}^{-1} \! dx \frac{(1 - x) K(x)}{x^2} - 2 K(0) + \frac{\pi}{2} Y(0)
\\
&- \pi Y'(0) - \! \int_0^1 \! da [K(a) - K(-a)]
\\
&+ \! \int_0^1 \! da \! \int_0^a \! db \frac{K'(b) - \! K(b) - \! K'(-b) + \! K(-b)}{b}. ~~~
\eeaa

\subsection{The low temperature limit: \texorpdfstring{$\sigma$}{sigma} for \texorpdfstring{$\gamma \ll 1$}{gamma<<1} and \texorpdfstring{$T \ll \omega_0$}{T<<1}}
\label{app:sigmalowT}

In order to find the simultaneous low-$\gamma$ and low-$T$ expansion of $\sigma$, we will combine our results above with the low-$T$ results obtained in Ref.~\cite{Hovhannisyan_2018}:
\beaa \label{sigmaiiGS}
\sigma_{11} &= \sigma_{11}^{(T=0)} + \frac{\pi}{3 \omega_0} \bigg[ \frac{T}{\omega_0} \bigg]^2 \gamma + \frac{1}{\omega_0} \circ \! \bigg( \bigg[ \frac{T}{\omega_0} \bigg]^2 \bigg),
\\
\sigma_{22} &= \sigma_{22}^{(T=0)} + \frac{2 \pi^2 \omega_0}{15} \bigg[ \frac{T}{\omega_0} \bigg]^4 \gamma + \omega_0 \circ \! \bigg( \bigg[ \frac{T}{\omega_0} \bigg]^4 \bigg), ~~~~~~
\eeaa
where
\beaa \label{sigmaiWC}
\sigma_{11}^{(T=0)} &= \frac{1}{\pi} \int_0^\infty d\omega \frac{J(\omega)}{|\alpha(\omega)|^2},
\\
\sigma_{22}^{(T=0)} &= \frac{1}{\pi} \int_0^\infty d\omega \frac{J(\omega)\omega^2}{|\alpha(\omega)|^2}.
\eeaa
We immediately notice that $\sigma_{11}^{(T=0)}$ and $\sigma_{22}^{(T=0)}$ are the same as, respectively, $\sigma_{11}$ and $\sigma_{22}$ but with $K(x)$ substituted by $F(x)$. Therefore, reading from Eqs.~\eqref{sgm11asympt}, \eqref{Phi}, \eqref{tashaxust}, \eqref{sgm22asympt}, and \eqref{Psi}
\beaa
\sigma_{11}^{(T=0)} &= \frac{1}{2 \omega_0} + \frac{\Phi_0}{2 \pi \omega_0} \gamma + \circ (\gamma),
\\
\sigma_{22}^{(T=0)} &= \frac{\omega_0}{2} + \frac{\omega_0 \Psi_0}{2 \pi} \gamma + \circ (\gamma),
\eeaa
where
\beaa \label{PhiPsi0}
\Phi_0 =& \int_{-\infty}^{-1} dx \frac{F(x)}{x^2} - 2 F(0) - \frac{\pi}{2} g(0) - \pi g'(0)
\\
&+ \int_0^1 da \int_0^a db \frac{F'(b) - F'(-b)}{b},
\\
\\
\Psi_0 =& \int_{-\infty}^{-1} dx \frac{(1 - x) F(x)}{x^2} - 2 F(0) - \pi g'(0)
\\
&+ \frac{\pi}{2} g(0) - \int_0^1 da [F(a) - F(-a)]
\\
&+ \int_0^1 da \int_0^a db \frac{F'(b) - F(b) - F'(-b) + F(-b)}{b}, ~~~~
\eeaa
with
\bea
g(0) = \frac{\omega_R^2 - \chi(\omega_0)}{\gamma \omega_0^2} \quad \mathrm{and} \quad g'(0) = \frac{\chi'(\omega_0)}{2 \gamma \omega_0}. ~~
\eea
With these, we can finally write:
\beaa
\sigma_{11} &= \frac{1}{2 \omega_0} + \frac{\Phi_0}{2 \pi \omega_0}\gamma + \frac{\pi}{3 \omega_0} \left[ \frac{T}{\omega_0} \right]^2 \! \gamma + \frac{1}{\omega_0} \alpha_2,
\\
\sigma_{22} &= \frac{\omega_0}{2} + \frac{\omega_0 \Psi_0}{2 \pi} \gamma + \frac{2 \pi^2 \omega_0}{15} \left[ \frac{T}{\omega_0} \right]^4 \! \gamma + \omega_0 \alpha_4, ~~~~
\eeaa
where
\bea
\alpha_m = \circ \bigg( \bigg[ 1 + \frac{T^m}{\omega_0^m}\bigg] \gamma \bigg);
\eea
in other words:
\beaa
\Phi_T &= \Phi_0 + \frac{2 \pi^2}{3} \bigg[ \frac{T}{\omega_0} \bigg]^2 + \circ \bigg( \bigg[ \frac{T}{\omega_0} \bigg]^2 \bigg),
\\
\Psi_T &= \Psi_0 + \frac{4 \pi^3}{15} \bigg[ \frac{T}{\omega_0} \bigg]^4 + \circ \bigg( \bigg[ \frac{T}{\omega_0} \bigg]^4 \bigg).
\eeaa
Lastly, since $T/\omega_0 \ll 1$, we will write the $\circ$ terms as $O \big( \gamma^{1 + \zeta_m} \big)$, with some $\zeta_m > 0$; note that $\zeta_2$ and $\zeta_4$ will generally be different.

In order to illustrate the above formulas on concrete examples, it is useful to write explicitly
\bea \nonumber
\chi(\omega) = \frac{\gamma \omega_0 \omega_c}{\pi} \bigg[ \int_0^\infty dz \frac{z f(z)}{z + \frac{\omega}{\omega_c}} + \mathcal{P} \int_0^\infty dz \frac{z f(z)}{z - \frac{\omega}{\omega_c}} \bigg].
\eea
Recalling Eqs.~\eqref{ggg} and \eqref{omerInt}, we also have
\beaa \nonumber
g(x) =& \frac{\sqrt{1 - x}}{\pi} \bigg[ \int_0^\infty dz \frac{f(z)}{z + \tom_0 \sqrt{1 - x}}
\\
&\hspace{2.2cm} - \mathcal{P} \int_0^\infty dz \frac{f(z)}{z - \tom_0 \sqrt{1 - x}} \bigg], ~~
\eeaa
where, for simplicity, we have introduced
\bea \label{bobol}
\tom_0 = \frac{\omega_0}{\omega_c}.
\eea

For a generic Lorentz-Drude spectral density, $J^{(\mathrm{L})}(\omega) = 2\gamma \omega_0 \omega \frac{\omega_c^2}{\omega_c^2 + \omega^2}$, the cutoff function will be (cf. Eq.~\eqref{fLor})
\bea
f^{(\mathrm{L})}(z) = \frac{2}{1 + z^2},
\eea
for which, it is a simple exercise to show that
\bea \label{chiL}
\chi^{(\mathrm{L})}(\omega) = \frac{2 \gamma \omega_0 \omega_c}{1 + \frac{\omega^2}{\omega_c^2}}
\eea
and
\bea
g^{(\mathrm{L})}(x) = \frac{2 \tom_0 (1 - x)}{1 + \tom_0^2 (1 - x)}.
\eea
Taking into account that
\bea
F^{(\mathrm{L})}(x) = \frac{2}{1 + \tom_0^2 (1 - x)},
\eea
it is also easy to calculate the integrals in Eq.~\eqref{PhiPsi0}, which brings us to
\beaa \label{PhiPsi0Lor}
\Phi^{(\mathrm{L})}_0 &= \frac{\pi \tom_0 (1 - \tom_0^2) - 2 (1 + \tom_0^2) + 4 \tom_0^2 \ln \tom_0}{(1 + \widetilde{\omega}_0)^2}, ~~~~
\\
\Psi^{(\mathrm{L})}_0 & = \frac{\pi \tom_0 (3 + \tom_0^2) - 2 (1 + \tom_0^2) - 4 \ln \tom_0}{(1 + \tom_0^2)^2}.
\eeaa

\subsection{The output ergotropy and connection-disconnection work for \texorpdfstring{$\gamma \ll 1$}{gamma<<1}}
\label{app:weakEW}

Let us now use Eqs.~\eqref{sgm11asympt} and \eqref{sgm22asympt} in Eqs.~\eqref{ergo:simp} and \eqref{wcd:eq} in order to obtain the corresponding expansions of $\mathcal{E}^\CL$ and $W_\cd^\CL$ with respect to $\gamma$. In order to do so, we first note that, since the high-temperature limit is already covered by Eqs.~\eqref{equipart:q} and \eqref{equipart:p}, the pertinent regime to explore is temperatures bounded from above by a constant \textit{not} $\gg \omega_0$. For such temperatures, $\coth \frac{\omega_0}{2 T}$ is a number close to $1$, therefore, $\gamma \ll 1$ is equivalent to $\gamma \ll \coth \frac{\omega_0}{2 T}$. Therefore, for $\mathcal{E}^\CL$, we can use the $\sqrt{a + x} = \sqrt{a} + \frac{x}{2 \sqrt{a}} + O (x^2)$, where $a > 0$ and $x \ll a$, expansion to straightforwardly arrive at
\bea
\frac{\mathcal{E}^\CL}{\omega_0 \gamma^2} = \frac{(\Phi_T - \Psi_T)^2}{16 \pi^2 \coth \frac{\omega_0}{2 T}} + \circ (1).
\eea

Turning to $W_\cd^\CL$, let us remind Eq.~\eqref{omerta}, namely, that
\bea
\frac{\omega_R^2}{\omega_0} = 2 \gamma \omega_c \hat{f},
\eea
where
\bea
\hat{f} = \int_0^\infty dz f(z)
\eea
is a number of the order of $1$ (it is exactly $1$ for Lorentz-Drude and exponential cutoff functions). Typically, $\omega_c$ is of the order of $\omega_0$, therefore, $\frac{\omega_R^2}{\omega_0^2} = O (\gamma)$. However, in certain situations, it may happen that $\omega_c \gg \omega_0$, so that $\gamma \omega_c \geq O (\omega_0)$, in which case $\omega_R$ will not be small anymore. Therefore, although $\gamma \omega^2_R / \omega_o^2$ will typically be $\circ (\gamma)$, we will keep the term proportional to in the expression for $W_\cd^\CL$. So,
\beaa
\frac{W_\cd^\CL}{\omega_0 \gamma} =& \; 2 \frac{\omega_c}{\omega_0} \hat{f} \coth \frac{\omega_0}{2 T} + \frac{\Phi_T - \Psi_T}{2 \pi}
\\
&+ \gamma \frac{\omega_c}{\omega_0} \hat{f} \, \frac{2 \Phi_T + \Psi_T}{2 \pi} + \circ (1).
\eeaa

In the low-temperature limit, which we described in Sec.~\ref{app:sigmalowT} above, for the formulas below, we will absorb the $O (\gamma \omega_c / \omega_0)$ and $O (T^4 / \omega_0^4)$ terms into the $\circ (1)$:
\beaa
\frac{\mathcal{E}^\CL}{\omega_0 \gamma^2} &= \frac{(\Phi_0 - \Psi_0)^2}{16 \pi^2} + \frac{\Phi_0 - \Psi_0}{12} \bigg[ \frac{T}{\omega_0} \bigg]^2 + \circ (1), ~~~~~~
\\
\frac{W_\cd^\CL}{\omega_0 \gamma} &= 2 \frac{\omega_c}{\omega_0} \hat{f} + \frac{\Phi_0 - \Psi_0}{2 \pi} + \frac{\pi}{3} \bigg[ \frac{T}{\omega_0} \bigg]^2 + \circ (1),
\eeaa
where $\Phi_0$ and $\Psi_0$ are given in Eq.~\eqref{PhiPsi0} and their values in the specific case of Lorentz-Drude cutoff are given in Eq.~\eqref{PhiPsi0Lor}.

\section{{THE SCALING OF THE COVARIANCE MATRIX IN THE ULTRASTRONG-COUPLING LIMIT}}
\label{app:strongsigma}

Let us now study $\sigma_{11}^\infty$ and $\sigma_{22}^\infty$ in the $\gamma \to \infty$ limit. Switching the integration variable in Eq.~\eqref{sgmii} to
\bea
\lom = \frac{\omega}{\omega_0 \sqrt{\gamma}},
\eea
we transform the expressions for $\sigma_{11}^\infty$ and $\sigma_{22}^\infty$ into
\beaa \label{zhiro}
\sigma_{11}^\infty &= \frac{1}{\omega_0 \sqrt{\gamma}} \frac{1}{\pi} \int_0^\infty d\lom \frac{\widetilde{f}_\gamma(\lom) \coth \frac{\lom \omega_0 \sqrt{\gamma}}{2 T}}{\big[ \lom^2 - \frac{1}{\gamma} - \widetilde{g}_\gamma(\lom) \big]^2 + \widetilde{f}_\gamma(\lom)^2},
\\
\sigma_{22}^\infty &= \omega_0 \sqrt{\gamma} \frac{1}{\pi} \int_0^\infty d\lom \frac{\widetilde{f}_\gamma(\lom) \, \lom^2 \coth \frac{\lom \omega_0 \sqrt{\gamma}}{2 T}}{\big[ \lom^2 - \frac{1}{\gamma} - \widetilde{g}_\gamma(\lom) \big]^2 + \widetilde{f}_\gamma(\lom)^2}, ~~~~~~
\eeaa
where we have introduced
\bea \label{paulaner}
\widetilde{f}_\gamma(\lom) := \sqrt{\gamma} \, \lom f\p{\tom_0 \lom \sqrt{\gamma}},
\eea
where the function $f$ is defined in Eq.~\eqref{J:def}, $\tom_0$ is given in Eq.~\eqref{bobol}, and $\widetilde{g}$, an analog of $g$, is defined as
\bea
\widetilde{g}_\gamma(\lom) = \frac{\omega_R^2 - \chi(\lom \omega_0 \sqrt{\gamma})}{\gamma \omega_0^2}.
\eea

Now, suppose that
\bea \label{cond1}
\lim_{z \to \infty} z f(z) = 0;
\eea
this is the case for the standard exponential ($f^{\mathrm{(exp)}}(z) = \pi e^{-z}$) and Lorentz-Drude ($f^{\mathrm{(L)}}(z) = 2/(1+z^2)$) cutoff functions. Note that this condition is essentially equivalent to requiring that $\omega_R^2 = 2 \omega_0 \omega_c \gamma \frac{1}{\pi} \int_0^\infty dz f(z)$ is finite.

Furthermore, it is also the case for the standard Lorentz-Drude and exponential cutoff functions that
\bea \label{cond2}
\exists \lim_{\lom \to \infty} \widetilde{g}_\gamma(\lom) := g_\infty < \infty,
\eea
and we will, in this section, work under this assumption.

Finally, since Eq.~\eqref{cond1} tells us that $\widetilde{f}_\gamma (\lom) \to 0$ as $\gamma \to \infty$, we recall the Dirac's delta function's representation in Eq.~\eqref{DDF} and use it in Eqs.~\eqref{zhiro} to conclude that, as $\gamma \to \infty$,
\beaa \nonumber
\sigma_{11}^\infty &\approx \frac{1}{\omega_0 \sqrt{\gamma}} \int_0^\infty d\lom \, \delta \bigg[ \lom^2 - \frac{1}{\gamma} - \widetilde{g}_\gamma(\lom) \bigg] \coth \frac{\lom \omega_0 \sqrt{\gamma}}{2 T},
\\
\sigma_{22}^\infty &\approx \omega_0 \sqrt{\gamma} \int_0^\infty d\lom \, \delta \bigg[ \lom^2 - \frac{1}{\gamma} - \widetilde{g}_\gamma(\lom) \bigg] \lom^2 \coth \frac{\lom \omega_0 \sqrt{\gamma}}{2 T}.
\eeaa
Here, since we are interested in only zero-order terms, we can neglect the $\frac{1}{\gamma}$ inside the $\delta$ functions, substitute the $\coth$'s with $1$ and $\widetilde{g}_\gamma (\lom)$ with $g_\infty$, according to Eq.~\eqref{cond2}. Hence,
\beaa \label{karush}
\sigma_{11}^\infty &\approx \frac{1}{\omega_0 \sqrt{\gamma}} \int_0^\infty d\lom \delta \p{\lom^2 - g_\infty} = \frac{1}{2 \omega_0 \sqrt{g_\infty \gamma}},
\\
\sigma_{22}^\infty &\approx \omega_0 \sqrt{\gamma} \int_0^\infty d\lom \delta \p{\lom^2 - g_\infty} \lom^2 = \frac{1}{2} \omega_0 \sqrt{g_\infty \gamma}. ~~~~~
\eeaa

Coming back to $g_\infty$, it is easy to see from Eq.~\eqref{chiL} that, for the Lorentz-Drude cutoff function,
\bea
\widetilde{g}^{\mathrm{(L)}}(\omega) = \frac{2 \omega_c}{\omega_0} \frac{\omega^2}{\omega^2 + \omega_c^2} \overset{\omega \to \infty}{\xrightarrow{\hspace*{12mm}}} \frac{2 \omega_c}{\omega_0} = g_\infty^{\mathrm{(L)}}. ~
\eea
For the exponential cutoff function, $f^{\mathrm{(exp)}}(z) = \pi e^{-z}$, it is a straightforward exercise to show, starting from the definition of $\chi$ (Eq.~\eqref{chichi}), that
\bea \label{mozhzh1}
\widetilde{g}^{\mathrm{(exp)}}(\omega) = \, \frac{2 \omega_c}{\omega_0} \mathcal{P} \int_0^\infty dt e^{-t} \frac{\omega^2}{\omega^2 - \omega_c^2 t^2} = \frac{2 \omega_c}{\omega_0} \, \Xi, ~~~~~
\eea
where
\beaa \label{mozhzh2}
\Xi =& \, \frac{\omega}{\omega_c} e^{-\frac{\omega}{\omega_c}} \bigg( \int_0^1 \frac{2 \, dt}{4 - t^2} \bigg[ \frac{2}{t} \sinh \frac{\omega t}{\omega_c} + \cosh \frac{\omega t}{\omega_c}\bigg] ~~~~~~
\\
& \hspace{1.7cm} - \int_1^\infty \frac{dt}{t} \frac{e^{- \frac{\omega t}{\omega_c}}}{2 + t} \bigg).
\eeaa
We immediately notice that, in the $\omega/\omega_c \to \infty$ limit, the second integral in Eq.~\eqref{mozhzh2} is damped at least as fast as $e^{-\frac{\omega}{\omega_c}}$. The first integral is clearly dominated by the values at $t = 1$. Therefore, switching the integration variable to $z = \frac{\omega}{\omega_c} (1 - t)$, we find that
\beaa \nonumber
\int_0^1 \frac{2 \, dt}{4 - t^2} & \bigg[ \frac{2}{t} \sinh \frac{\omega t}{\omega_c} + \cosh \frac{\omega t}{\omega_c} \bigg]
\\
=& \, e^{\frac{\omega}{\omega_c}} \frac{\omega_c}{\omega} \int_0^{\frac{\omega}{\omega_c}} dz e^{-z} \bigg[ 1 + \frac{z^2 \omega_c^2}{\omega^2} + O \bigg( \frac{z^3 \omega_c^3}{\omega^3}\bigg) \bigg],
\eeaa
which means that
\bea \nonumber
\int_0^1 \frac{2 \, dt}{4 - t^2} \bigg[ \frac{2}{t} \sinh \frac{\omega t}{\omega_c} + \cosh \frac{\omega t}{\omega_c} \bigg] = e^{\frac{\omega}{\omega_c}} \frac{\omega_c}{\omega} \bigg[1 + O \bigg( \frac{\omega_c^2}{\omega^2} \bigg) \bigg].
\eea
Plugging this into Eq.~\eqref{mozhzh2}, we finally find that
\bea
\Xi = 1 + O \bigg( \frac{\omega_c^2}{\omega^2} \bigg);
\eea
in other words,
\bea
g_\infty^{\mathrm{(exp)}} = \lim_{\omega \to \infty} \widetilde{g}^{\mathrm{(exp)}}(\omega) = \frac{2 \omega_c}{\omega_0} \quad \big( = g_\infty^{\mathrm{(L)}}\big).
\eea
Lastly, let us mention that, since the asymptotic analysis here relies on $\tom_0 \lom \sqrt{\gamma} \gg 1$ (cf. Eqs.~\eqref{paulaner} and \eqref{cond1}), the $\gamma \to \infty$ limit is to be understood as
\bea
\gamma \gg \bigg( \frac{\omega_c}{\omega_0}\bigg)^2.
\eea
We would also like to emphasize that the scalings in Eq.~\eqref{karush} are not limited to Lorentz-Drude and exponential cutoff functions: they hold whenever the conditions \eqref{cond1} and \eqref{cond2} are satisfied (although note that the scalings in Eq.~\eqref{karush} will hold also when the condition \eqref{cond2} is weakened to $0 < \lowlim\limits_{\omega \to \infty} \widetilde{g}(\omega) \leq \uplim\limits_{\omega \to \infty} \widetilde{g}(\omega) < \infty$), which, we believe, is what happens generically.

\bibliographystyle{prrK}

\bibliography{references}

\end{document}